\documentclass[seceq]{ptptex}
\usepackage{amsmath,amssymb}

\newcommand{\disp}{\displaystyle}
\newcommand{\pr}{\prime}

\newcommand{\pd}{\partial}
\newcommand{\ok}{\{\!\!\{}
\newcommand{\ck}{\}\!\!\}}

\def\bt{\breve{t}}
\def\wtlam{\widetilde{\lambda}}
\def\wtPhi{\widetilde{\Phi}}
\def\wtbPhi{\widetilde{\breve{\Phi}}}
\def\boldv{\mbox {\boldmath{$v$}}}
\def\m2{\hspace*{-0.2cm}}

\def\gm{\mbox {$\gamma$}}

\def\th{\theta}
\def\vth{\vartheta}
\def\balph{\mbox {$\breve{\alpha}$}}
\def\bbeta{\mbox {$\breve{\beta}$}}
\def\btau{\mbox {$\breve{\tau}$}}

\def\bg{\mbox {$\breve{g}$}}
\def\bm{\mbox {$\breve{m}$}}

\def\by{\mbox {$\breve{y}$}}

\def\balam{\mbox {$\bar{\lambda}$}}
\def\brlam{\mbox {$\breve{\lambda}$}}
\def\brLam{\mbox {$\breve{\Lambda}$}}

\def\bphi{\mbox {$\breve{\phi}$}}
\def\bPhi{\mbox {$\breve{\Phi}$}}

\def\bth{\mbox {$\breve{\theta}$}}
\def\bvth{\mbox {$\breve{\vartheta}$}}

\def\bxi{\mbox {$\breve{\xi}$}}
\def\tbxi{\mbox {\tiny $\breve{\xi}$}}
\def\bA{\breve{A}}

\def\bD{\breve{D}}
\def\bF{\breve{F}}
\def\bG{\breve{G}}

\def\bv{\breve{v}}
\def\bW{\breve{W}}
\def\bY{\breve{Y}}
\def\bZ{\breve{Z}}
\def\bcA{\breve{\cal A}}
\def\cD{{\cal D}}

\def\cL{{\cal L}}
\def\cM{{\cal M}}
\def\dlam{\dot{\lambda}}
\def\half{{1 \over 2}}
\def\quarter{\frac{1}{4}}
\def\sqr2{\sqrt{2}}
\def\rTr{\rm Tr}
\def\uPhi{\underline{\Phi}}
\def\uPsi{\underline{\Psi}}

\title{%        %You can use \\ for explicit line-break
Gauge Field Theory of Horizontal Symmetry Generated by\\
a Central Extension of the Pauli Algebra
}

\author{%       %Use \sc for the family name
Ikuo S. {\sc Sogami}\footnote{E-mail address: sogami@cc.kyoto-su.ac.jp}
}

\inst{%         %Affiliation, neglected when [addenda] or [errata]
Physics Department, Kyoto Sangyo University, Kyoto 603-8555, Japan
}

\recdate{%      %Editorial Office will fill in this.
\today
}

\abst{%         %this abstract is neglected when [addenda] or [errata]
The standard model of particle physics is generalized so as to be
furnished with a horizontal symmetry generated by an intermediary algebra
between simple Lie algebras $\mathfrak{su}(2)$ and $\mathfrak{su}(3)$.
Above a certain high energy scale $\brLam$, the horizontal gauge symmetry
is postulated to hold so that the basic fermions, quarks and leptons,
form its fundamental triplets, and a triplet and singlet of the horizontal
gauge fields distinguish generational degrees of freedom. A horizontal
scalar triplet is introduced to make the gauge fields super-massive
by breaking the horizontal symmetry at $\brLam$. From this scalar triplet,
there emerge real scalar fields which do not interact with fermions except
for neutrino species and may give substantial influence on evolution of
the universe. Another horizontal scalar triplet which breaks the electroweak
symmetry at a low energy scale $\Lambda\simeq 2\times 10^2$GeV reproduces
all of the results of the Weinberg-Salam theory, produces hierarchical
mass matrices with less numbers of unknown parameters in a unified way
and predicts six massive scalar particles, some of which might be
observed by the future LHC experiment. 
}

\begin{document}

\maketitle

\section{Introduction}
The Standard Model of particle physics (SM) still possesses a vast area
being full of unsettled problems. It is not clear why quarks and leptons
exist in three generations\cite{KobayashiMaskawa} and why they
possess characteristic mass spectra with hierarchical structure.
The SM has long been suffering the plethora of unknown coupling constants
in the Yukawa interaction. To investigate systematically this uncultivated
fertile area, which goes customarily by the name of the flavor physics,
we develop an exploratory scheme of gauge field theory of a horizontal (H)
symmetry in this paper.

%Search for such a H symmetry in complicated low energy phenomena of flavor
%physics is necessarily not a facile task. A genuine H symmetry could only
%be discovered after repeated trial and error. It seems a permissible attempt
%to postulate that a H symmetry is realized by a Lie group generated by
%an intermediary algebra between the simple Lie algebras $\mathfrak{su}(2)$
%and $\mathfrak{su}(3)$.

As a continuous group for a H symmetry, we adopt here a Lie group $\bG(\bcA)$ 
generated by a central extension of the Pauli algebra, $\bcA$, which was
found in analysis of quark mass spectra and applied to characterize
the flavor mixing matrices (FMMs)\cite{KobayashiMaskawa,Cabibbo} in
the previous papers.~\cite{Sogami,SogamiKon}
While the Pauli algebra consists of three independent $2 \times 2$ matrices,
the new algebra $\bcA$ which is an intermediary algebra between 
$\mathfrak{su(2)}$ and $\mathfrak{su(3)}$ is composed of four independent
$3 \times 3$ matrices.
%We interpret a Lie group $\bG(\bcA)$ generated on
%the algebra $\bcA$ as a group of hitherto-unrevealed H symmetry, $\H$,
%which governs a basic framework of flavor physics.
%A Lie group $\bG(\bcA)$ generated on the algebra $\bcA$ was used,
%{\em restrictively}, in the previous works for analyses of accumulated
%data of the flavor mixing matirices (FMMs). Here, we abandon such
%a restricted interpretation and adopt a viewpoint that the Lie group
%$\bG(\bcA)$ embodies a group of hitherto-unrevealed H symmetry, $\H$,
%which determines a basic framework of flavor physics.

Successes of the SM in low energy physics naturally require that
our gauge theory inherits its contents such as the basic fermion fields
and the gauge fields for the vertical (V) symmetry. To formulate a simple
generalization of the SM with the H symmetry, we add a minimal set of
gauge and scalar fields. The theory necessitates a triplet and singlet
of gauge fields of the H symmetry, which correspond to four independent
generators of the algebra $\bcA$. In place of the electroweak (EW) scalar
doublet in the SM, two types of scalar triplets of the H symmetry,
$\bPhi(x)$ and $\Phi(x)$, are postulated to exist in the present formalism.

The H and EW symmetries are broken at two stages with high and low energy
scales, $\brLam$ and $\Lambda$\, (\,$\brLam\gg\Lambda$\,).~\footnote{This
assumption brings in another sort of the hierarchy problem. We will
not deal with it, here, in expectation of future resolution of
the celebrated {\lq\lq}Hierarchy{\rq\rq} puzzle\cite{Zee}. One possible
resolution is to accept the supersymmetric extension.} Here we develop
a new compound mechanism of symmetry breakdown so as to formulate
an effective theory with less numbers of unknown parameters for
low energy phenomenology. First, a scale of partial breakdown of
symmetry is fixed tentatively by finding a local stationary point of
the Higgs potential for the triplet $\bPhi(x)$ ($\Phi(x)$). Then, we
make full use of possible freedoms in the decomposition of the triplet
$\bPhi(x)$ ($\Phi(x)$) so that the residual symmetry is broken and
unphysical modes of component fields are forbidden to appear.

At the high energy scale $\brLam$, the triplet $\bPhi(x)$ breaks the H
symmetry so that all of the gauge fields are transformed into super-massive
vector fields and the neutrinos acquire Majorana masses. From this triplet,
there emerge the scalar fields that interact solely with neutrino species.
Those fields could have substantial influences on the evolution of
the universe.
%
%If some of those fields has a smaller mass than those of all
%other scalar fields produced at the high and low energy scales,
%
%Such a particle might be related to the dark mass or the dark
%energy.\cite{Riess1,Perlmutter,Riess2,WMAP}

The horizontal symmetry works to reduce the number of the Yukawa
coupling constants down to 4/9 of those of the SM.
The triplet $\Phi(x)$ breaking the EW symmetry at the low energy scale,
$\Lambda \simeq \Lambda_{\rm EW} = 10^{2}$\,GeV, reproduces all results
of the Weinberg-Salam theory and produces mass matrices of Dirac type
which possess the same number of unknown parameters with the Yukawa
coupling constants and create hierarchical mass spectra.
The breakdown of the EW symmetry results in six different types of
massive scalar fields. It turns out possible to make all of their
masses so large that the flavor changing neutral currents (FCNCs) 
can be suppressed.

In \S 2, the algebra $\bcA$ and associated group $\bG(\bcA)$ are reconsidered
in some detail.~\cite{Sogami} Contents of bosonic fields being necessary
to formulate a gauge field theory of the V and H symmetries are examined
in \S 3. We construct explicitly the Lagrangian density of the theory out
of invariants of these symmetries in \S 4. In \S 5, remark is given on
how to use possible freedoms in the decompositions of the scalar triplets
so that unphysical modes are effectively forbidden to appear in the phases
of the broken symmetries. Breakdowns of the H symmetry at the high energy
scale $\brLam$ and the EW symmetry at the low energy scale $\Lambda$ are
investigated, respectively, in \S 6 and \S 7. We discuss general remarks
of our formalism and point out open problems in \S 8. Analysis on identical
equations among EW and H invariants of the scalar triplets is made in Appendix.

\section{Horizontal symmetry}
%To account for a rich variety of phenomena in the low energy flavor physics,
%the V and H symmetries and their breaking mechanisms must have flexible and
%elaborate structure. The central extension of the Pauli algebra $\bcA$ and
%associated Lie group $\bG(\bcA)$\footnote{The group $\bG(\bcA)$ which is
%generated by linearly independent elements extracted from $3 \times 3$
%matrix representations of the discrete $S_3$ symmetry\cite{Sogami} can be
%interpreted as a continuous quantum extension of the discrete classical
%group $S_3$.} enable us to formulate properly the gauge field theory
%with such characteristics.
%
%\subsection{Central extension of the Pauli algebra}
The central extension of the Pauli algebra, $\bcA$, is a closed subalgebra
of $\mathfrak{su(3)}$. Four independent generators of the algebra
$\bcA$ are formed by linear combinations of the Gell-Mann matrices
$\lambda_j\ (j=1,\,2,\cdots,\,8)$ for $\mathfrak{su(3)}$. Its center
is the projective element defined by
\begin{equation}
    \bD\ =\ \frac{1}{3}(I+\lambda_1+\lambda_4+\lambda_6)
     \ = \ 
      \frac{1}{3}
      \begin{small}
        \left(
         \begin{array}{ccc}
           1 & 1 & 1\\
           1 & 1 & 1\\
           1 & 1 & 1\\
         \end{array}
        \right)
      \end{small}
\end{equation}
which possesses the idempotent property $\bD^2 = \bD$. This is the element
known as a {\em democratic matrix}~\cite{DemocraticMM} which acts to provide
the quarks and charged leptons with hierarchical mass spectra. Three other
linearly independent generators of the algebra are given as follows:
\begin{equation}
 \left\{
   \begin{array}{lclcl}
    \btau_1 \!\! &=& \frac{1}{\sqrt{3}}(\lambda_3-\lambda_4+\lambda_6)
            &=& \!\!\!\frac{1}{\sqrt{3}}
            \begin{small}
            \left(
             \begin{array}{rrr}
              1\! &   0\! &  -1\\
             \noalign{\vskip 0.05cm}
              0   &  -1   &   1\\
             \noalign{\vskip 0.05cm}
             -1   &   1   &   0\\
             \end{array}
            \right)
             \end{small}
            \;,\\
   \noalign{\vskip 0.2cm}
    \btau_2 \!\! &=&
           \frac{1}{\sqrt{3}}(\lambda_2-\lambda_5+\lambda_7)
           &=&\!\!\!\frac{1}{\sqrt{3}}
           \begin{small}
           \left(
             \begin{array}{rrr}
              0\! &  -i\! &  i\\
             \noalign{\vskip 0.05cm}
              i   &   0   & -i\\
             \noalign{\vskip 0.05cm}
             -i   &   i   &   0\\
             \end{array}\,\,
            \right)
            \end{small}
              \;,\\
   \noalign{\vskip 0.2cm}
    \btau_3  \,\, &=& 
            \frac{1}{3}(-2\lambda_1+\lambda_4+\lambda_6+\sqrt{3}\lambda_8)
            &=& \, \frac{1}{3}
            \begin{small}
            \left(
             \begin{array}{rrr}
              1\! &  -2\! &   1\\
             \noalign{\vskip 0.05cm}
             -2   &   1   &   1\\
             \noalign{\vskip 0.05cm}
              1   &   1   &  -2\\
             \end{array}\,
            \right)
            \end{small}
            \;,
  \end{array}\
 \right.
 \label{SU(2)basis}
\end{equation}
which obey the Pauli-type product rules
\begin{equation}
   \btau_j\,\btau_k = \delta_{jk}\,(I-\bD) + i\,\epsilon_{jkl}\,\btau_l
   \label{Paulilike}
\end{equation}
and the normalization condition $\rTr(\btau_j\btau_k) = 2\delta_{jk}$.
These elements and the democratic element $\bD$ are orthogonal in
the sense that $\bD\,\btau_j = \btau_j\,\bD =0$. Here, we identify
the Lie algebra which generates the basic H symmetry for the flavor
physics with the central extension of the Pauli algebra constructed as
\begin{equation}
    \bcA = \{ \bD,\,\btau_1,\,\btau_2,\,\btau_3 \}.
\end{equation}

The Lie group $\bG(\bcA)$\footnote{The group $\bG(\bcA)$ which is
generated by linearly independent elements extracted from $3 \times 3$
matrix representations of the discrete $S_3$ symmetry\cite{Sogami} can be
interpreted as a continuous quantum extension of the discrete classical
group $S_3$.} is defined by the set of exponential mappings of
all possible linear combinations of the elements of the algebra $\bcA$ as
follows:
\begin{equation}
 \hspace*{-0.4cm}
   \bG(\bcA) = \left\{\Omega(\vth)=\exp{\left( \disp i\vth_0 \bD
             + i\sum_{j=1}^3\,\vth_j\,\btau_j\right)}:
             \vth_0, \vth_j \in \mathbb{R} \right\}\;.
 \label{Liegroup}
\end{equation}
This Lie group $\bG(\bcA)$ which is postulated to generate the H symmetry
group possesses two subgroups
\begin{equation}
 SU_{\rm H}(2)=\left\{\Omega_2(\vth)=
                       \exp{\left(\disp i\sum_{j=1}^3\,\vth_j\,\btau_j\right)}
             :\,\vth_j \in \mathbb{R} \right\}
 \label{groupelement}
\end{equation}
and
\begin{equation}
   U_{\rm H}(1) = \left\{\,\Omega_1(\vth_0)=\exp({i\vth_0 \bD}):\, \vth_0 
                 \in \mathbb{R}\,\right\} .
\end{equation}
For analyses below, it is convenient to use the following expression of
the group elements in terms of the elements of the algebra $\bcA$ as
\begin{equation}
 \Omega_2(\vth)%=\exp{\left(\disp i\sum_{j=1}^3\,\vth_j\,\btau_j\right)}
 = \bD + \cos\Theta(I-\bD) + \frac{\sin\Theta}{\Theta}
   \left(i\sum_{j=1}^3\,\vth_j\btau^j\right),\ \ 
 \Omega_1(\vth_0)%=\exp({i\vth_0 \bD})
                  = e^{i\vth_0}\bD + (I-\bD)
 \label{GroupElementinAlgebra}
\end{equation}
where $\Theta^2 = \sum_{j=1}^3\,\vth_j^2$.

%The H symmetry thus constructed, $\bG(\bA)=SU_{\rm H}(2)\times U_{\rm H}(1)$,
%is isomorphic to the WS group $SU_L(2)\times U(1)$ of EW interaction
%at the group level. However, their representations and actions on quantum
%states are totally different. While the latter operates on the two
%dimensional spinor multiplets, the former induces transformations
%among three dimensional spinor multiplets.

%To derive invariant quantities of the H symmetry, it is necessary
%to examine the behaviors of the group elements under the operations of
%complex conjugate and transposition. 
From explicit representations of the matrices $\btau_j$ in (\ref{SU(2)basis}),
we obtain the relation of complex conjugation,
$\btau_2(i\btau_j)^{\ast}=(i\btau_j)\btau_2$, which leads readily to
$\btau_2\,(\Omega_2(\vth))^{\ast} = \Omega_2(\vth)\,\btau_2$.
As for the transposition, care must be taken to distinguish matrices
and multiplets of the H symmetry from those of the EW and Lorentz
symmetries\footnote{For simplicity, multiplet structure for the color
symmetry is not shown explicitly in this paper.}. Here, in addition to
the ordinary symbol $t$ which is used to transpose the quantities
related with the EW and the Lorentz degrees of freedom, a new
symbol $\bt$ is introduced for the transposition operation for the
H symmetric degrees of freedom. From (\ref{SU(2)basis}), one can prove
the relation $\btau_2\,{}^{\bt}(i\btau_j)=(-i\btau_j)\btau_2$
which results in the identity
%\begin{equation}
% \btau_2\,\raisebox{0.1cm}{$\bt$}
%           \{\exp{(i\sum_{j=1}^3\,\vth_j\,\btau_j)}
%           \}
%         = \{\exp{(-i\sum_{j=1}^3\,\vth_j\,\btau_j)}
%           \} \btau_2
%\end{equation}
$\btau_2\,{}^{\bt}\Omega_2(\vth) = \Omega_2(-\vth)\,\btau_2
                                 = (\Omega_2(\vth))^{-1}\,\btau_2$.
Note that the Pauli matrices $\tau_j$ and the matrices $\btau_j$ in
(\ref{SU(2)basis}) have the same behaviors under the operations of
complex conjugate and transposition.

%\subsection{Kernels of $\bD$ and $(\btau_1,\ \btau_2,\ \btau_3)$}
The simultaneous eigenstates of the generators $\btau_3$ and $\bD$ are
found in the forms
\begin{equation}
  \begin{array}{lll}
      |\,1\ \rangle
        = \frac{1}{\sqrt{2}}\,\raisebox{0.2cm}{}^{\bt}(1,\,-1,\,0),
     &
       |\,2\ \rangle
         = \frac{1}{\sqrt{6}}\,\raisebox{0.2cm}{}^{\bt}(1,\,1,\,-2),
     &
       |\,3\ \rangle
         = \frac{1}{\sqrt{3}}\,\raisebox{0.2cm}{}^{\bt}(1,\,1,\,1).
  \end{array}
  \label{eigenvectors}
\end{equation}
Note that the doublet $\{\,|\,1\ \rangle,\, |\,2\ \rangle\,\}$
and singlet $\{\,|\,3\ \rangle\,\}$ are irreducible representations
of the discrete permutation symmetry $S_3$.
These eigenstates form a convenient basis to represent fundamental
multiplets of the H symmetry.
%Actions of the generators $\btau_j$ and $\bD$
%on these vectors are summarized as follows:\cite{Sogami}
%\begin{equation}
% \begin{array}{lll}
%  \btau_1|\,1\ \rangle=|\,2\ \rangle, &
%  \btau_1|\,2\ \rangle=|\,1\ \rangle, &
%  \btau_1|\,3\ \rangle=0,\\
%  \noalign{\vskip 0.3cm}
%  \btau_2|\,1\ \rangle=i|\,2\ \rangle, &
%  \btau_2|\,2\ \rangle=-i|\,1\ \rangle, &
%  \btau_2|\,3\ \rangle=0,\\
%  \noalign{\vskip 0.3cm}
%  \btau_3|\,1\ \rangle=|\,1\ \rangle, &
%  \btau_3|\,2\ \rangle=-|\,2\ \rangle, &
%  \btau_3|\,3\ \rangle=0,\\
%  \noalign{\vskip 0.3cm}
%  \bD|\,1\ \rangle=0, &
%  \bD|\,2\ \rangle=0, &
%  \bD|\,3\ \rangle=|\,3\ \rangle.\\
% \end{array}
%\end{equation}
Evidently, $\{|\,1\ \rangle, |\,2\ \rangle\}$ and $\{|\,3\,\rangle\}$
are, respectively, the kernels of the generator $\bD$ and
the set of the generators $\btau_1$, $\btau_2$ and $\btau_3$.
We utilize, in later argument, the fact that the state $|\,3\,\rangle$
is an eigenstate for all of the elements of the $\bG(\bA)$ group as
%\begin{equation}
%  \Omega(\vth)\,|\,3\,\rangle
%  = \exp{\left( \disp i\vth_0 \bD
%  + i\sum_{j=1}^3\,\vth_j\,\btau_j\right)}\,|\,3\,\rangle
%  = e^{i\vth_0}\,|\,3\,\rangle
%  \label{property3vector}
%\end{equation}
\begin{equation}
  \Omega(\vth)\,|\,3\,\rangle
  = \Omega_2(\vth)\Omega_1(\vth_0)\,|\,3\,\rangle
  = e^{i\vth_0}\,|\,3\,\rangle
  \label{property3vector}
\end{equation}
where $\vth_j$ $(j=0,\cdots,3)$ are arbitrary real numbers.

To illustrate a unique feature of the H symmetry, let us consider here
an arbitrary H triplet $T (=|\,T\,\rangle )$ and introduce an operation
$\ok \cdots \ck$ on $T$ by
\begin{equation}
\ok T \ck = \sum_{i=1}^3\,T_i = \sqrt{3}\,\langle\, 3\,|\,T\,\rangle .
\end{equation}
Then, owing to the property of the vector $|\,3\,\rangle$ in
(\ref{property3vector}), the group action $T \rightarrow \Omega(\vth)\,T$
induces the transformation
$\ok T\ck \rightarrow \ok\Omega(\vth)\,T\ck = e^{i\vth_0}\ok T\ck$
on $\ok T\ck$. Therefore, the quantity $\ok T \ck$, which we call
hereafter the H-sum of $T$, behaves as an eigenvector for an arbitrary
elements of the $\bG(\bA)$ group. In particular, the H-sum is invariant
under the action of the $SU_{\rm H}(2)$ group, i.e.,
$\ok\Omega_2(\vth)\,T\ck = \ok T\ck$.

\section{Field contents in the gauge field theory of V$\times$H symmetry}
%In order to reproduce all results of the SM, our gauge theory must
%inherit, with appropriate modifications, all of its field contents
%such as the basic fermionic fields and the gauge fields for
%the V symmetry. In order to realize the simplest generalization of
%the SM with the H symmetry, we add new minimal sets of H multiplets of
%the gauge and scalar fields.
In the high energy region, three generations of fermions characterized by
the same quantum numbers of the V symmetry and definite chiralities
are postulated generically to form the fundamental H triplets as 
\begin{equation}
\Psi_h^{\,f}(x) = 
{}^{\bt}\left(\,\psi_{{h}1}^{\,f}(x),\,\psi_{{h}2}^{\,f}(x),\,
                \psi_{{h}3}^{\,f}(x)\,\right)
   \label{chiraltriplet}
\end{equation}
where $f\, (= q, u, d; l, \nu, e)$ distinguishes EW multiplets and
$h\, (= L, R)$ refers to chiral components. Specifically, $\Psi_L^{\,f}$
(=$\Psi_L^{\,q},\,\Psi_L^{\,l}$) is the H triplet of EW doublets
and $\Psi_R^{\,f}$ (=$\Psi_R^{\,u},\,\Psi_R^{\,d},\,\Psi_R^{\,\nu},
\,\Psi_R^{\,e}$) is that of EW singlets.

We represent the gauge fields for $SU_{\rm c}(3)$, $SU_L(2)$ and $U(1)$
subgroups of the V symmetry by $A_\mu^{(3)j}(x)\ (j=1,\cdots,8)$,
$A_\mu^{(2)j}(x)\ (j=1, 2, 3)$ and $A_\mu^{(1)}(x)$, respectively,
and write the gauge coupling constants for $A^{(k)}_\mu(x)$ by
$g_k\ (k=1,\,2,\,3)$.
%As usual, the field strengths of the gauge
%fields $A^{(k)}_\mu(x)$ are expressed by $F_{\mu\nu}^{(k)}$.
%
Corresponding to the generators $\{\,\btau_1,\,\btau_2,\,\btau_3\,\}$ and
$\{\,\bD\,\}$ of the H symmetry, a triplet $\bA^{(2)j}_\mu(x)$ ($j=1,\,2,\,3$)
and a singlet $\bA^{(1)}_\mu(x)$ of the gauge fields are postulated to exist
as follows:
\begin{equation}
      \btau_j\ \leftrightarrow\ \bA^{(2)j}_\mu(x),\quad
      \bD\ \leftrightarrow\ \bA^{(1)}_\mu(x)\; .
\end{equation}
The gauge fields $\bA^{(k)}_\mu(x)$ with coupling constants $\bg_k$
have the field strengths
\begin{equation}
     \bF_{\mu\nu}^{(2)j} = \pd_\mu\bA^{(2)j}_\nu - \pd_\nu\bA^{(2)j}_\mu
                  +\bg_2\,\epsilon_{jkl}\,\bA^{(2)k}_\mu\bA^{(2)l}_\nu,\ \ 
     \bF_{\mu\nu}^{(1)} = \pd_\mu\bA^{(1)}_\nu - \pd_\nu\bA^{(1)}_\mu .
 \label{curvatureH}
\end{equation}
The gauge fields $A_\mu^{(k)}(x)\ (k=1, 2, 3)$ and
$\bA_\mu^{(k)}(x)\ (k=1, 2)$ of the V and H symmetries belong, respectively,
to the singlet representations of the H and V symmetries.

To break properly the H and EM symmetries, two kinds of H multiplets of
scalar fields are presumed to exist. For the H symmetry breaking around
the high energy scale $\brLam$, we introduce three scalar fields
$\bphi_j(x)$ which belong to the V singlet and form the fundamental
triplet of the H symmetry as
\begin{equation}
 \bPhi(x) ={}^{\bt} \left(\,\bphi_1(x),\,\bphi_2(x),\,\bphi_3(x)\,\right)\;.
 \label{scalartriplet}
\end{equation}
This scalar triplet does not couple with the fermion fields
except for the right-handed neutrino triplet $\Psi_{\rm R}^{\nu}(x)$.
It is this character of $\bPhi(x)$ that prohibits
the fermion fields to acquire Dirac masses of the scale $\brLam$.
Note that a new triplet defined by
\begin{equation}
   \wtbPhi(x)=i\btau_2\bPhi^\ast(x)
   \label{AssociatebPhi}
\end{equation}
has the same transformation property with $\bPhi(x)$ under the action of
$SU_{\rm H}(2)$ group. We call this triplet as an associated triplet of
$\bPhi(x)$.

To form the Yukawa interaction and break it at the scale $\Lambda$,
a new set of scalar fields belonging to non-trivial multiplets of
both groups of EW and H symmetries is required to exist.
As a possible choice of such fields, we assume here that three EW
doublets $\Phi_j(x)$ with the EW hypercharge $Y_{\rm EW} = 1$
constitute the H triplet as
\begin{equation}
\begin{array}{lcl}
 \Phi(x) &=& {}^{\bt}\left(\,\Phi_1(x),\,\Phi_2(x),\,\Phi_3(x)\,\right)\\
         \noalign{\vskip 0.35cm}
         &=& \raisebox{0.5cm}{}^{\bt}\!\left(
            \begin{array}{ccc}
              \left(
                 \begin{array}{c}
                 \phi_1^+(x)\\
                 \noalign{\vskip 0.2cm}
                 \phi_1^0(x)\\
                 \end{array}
                \right),
                &
                \left(
                 \begin{array}{c}
                 \phi_2^+(x)\\
                 \noalign{\vskip 0.2cm}
                 \phi_2^0(x)\\
                 \end{array}
               \right),
                &
               \left(
                 \begin{array}{c}
                 \phi_3^+(x)\\
                 \noalign{\vskip 0.1cm}
                 \phi_3^0(x)\\
                 \end{array}
               \right)
             \end{array}
          \right)\ ,
\end{array}
\end{equation}
and define its associated triplet by
\begin{equation}
\begin{array}{lcl}
 \wtPhi(x) &=& (i\btau_2)(i\tau_2)\,\Phi^{\ast}(x)
      = (i\btau_2)\,{}^{\bt}\!\left(\,i\tau_2\Phi_1^\ast(x),\,
           i\tau_2\Phi_2^\ast(x),\,i\tau_2\Phi_3^\ast(x)\,\right)\\
           \noalign{\vskip 0.35cm}
           &=& (i\btau_2)\,\raisebox{0.5cm}{}^{\bt}\!\left(\,
              \left(
                 \begin{array}{c}
                 \phi_1^{0\ast}(x)\\
                 \noalign{\vskip 0.2cm}
                 -\phi_1^{-}(x)\\
                 \end{array}
                \right),\,\,
                \left(
                 \begin{array}{c}
                  \phi_2^{0\ast}(x)\\
                 \noalign{\vskip 0.2cm}
                 -\phi_2^{-}(x)\\
                 \end{array}
               \right),\,\,
               \left(
                 \begin{array}{c}
                 \phi_3^{0\ast}(x)\\
                 \noalign{\vskip 0.2cm}
                 -\phi_3^{-}(x)\\
                 \end{array}
               \right)
          \right) .
\end{array}
  \label{AssociatePhi}
\end{equation}
Both of these triplets have the same transformation properties under
the group of H symmetry, $SU_{\rm H}(2)$, and the EW isospin group,
$SU_{\rm EW}(2)$. The H triplets $\wtPhi(x)$ and $\Phi(x)$ interact,
respectively, with the right-handed fermion triplets of EW
up- and down-sectors, as in (\ref{YukawaUp}) and(\ref{YukawaDown}).

%All of the field contents which are postulated to exist in the present
%theory are listed in the Table 1. The H hypercharge $\by$ is entered
%in the last column.
It is postulated that the scalar triplet $\bPhi(x)$
possesses a non-vanishing H hypercharge ($\by_{\small \bPhi}\,\neq 0)$
and that all other triplets have zero H hypercharge.
Evidently, the H-sums $\ok \Psi^f_L(x) \ck$ and $\ok \Phi(x) \ck$
are EW doublets and $\ok \Psi^f_R(x) \ck$ are singlets.
The H-sum $\ok \bPhi(x) \ck$ carrying non-vanishing H hypercharge is
invariant under the H and EW symmetries.

%\begin{table}[t]
%\newcommand{\lw}[1]{\smash{\lower2.0ex\hbox{#1}}}
%\begin{center}
%\renewcommand{\arraystretch}{1.75}
%\caption{Field Contents of the New Gauge Theory}
%\smallskip
% \begin{tabular}{|c|c|c|c|c|}
%  \hline
%   Fields\         &\qquad Operators\qquad\quad
%                        &\ $SU(3)\times SU(2)\times U(1)$
%                        &\ \ \ $SU_{\rm H}(2)$\ \ \qquad&\by\\
%  \hline
%   \lw{Fermion fields}\ &\ $\Psi_L^{\,f}(x)$\quad 
%                        &\ (\,3,\ 2,\ 1\,),\ (\,1,\ 2,\,1\,)
%                         \qquad &\ 3\qquad&0\\
%  \cline{2-5}
%                        &\ $\Psi_R^{\,f}(x)$\quad 
%                        &\ (\,3,\ 1,\ 1\,),\ (\,1,\ 1,\,1\,)
%                         \qquad &\ 3\qquad&0\\
%  \hline
%   \lw{Gauge fields}\   &\ $A_\mu^{(k)}(x)$\quad
%                        &\ (\,8,\,1,\,1\,),\ (\,1,\,3,\,1\,),\ (\,1,\,1,\,1\,)
%                         \qquad &\ 1\qquad&0\\
%  \cline{2-5}
%                        &\ $\bA_\mu^a(x)$\quad
%                        &\ (\,1,\ 1,\ 1\,)\qquad &\ 3,\ 1\qquad&0\\
%  \hline
%  \lw{Scalar fields}    &\ $\bPhi(x)$\quad
%                        &\ (\,1,\ 1,\ 1\,)\qquad&\ 3\qquad&$\by_{\bPhi}$\\
%  \cline{2-5}
%                        &\ $\Phi(x)$\quad
%                        &\ (\,1,\ 2,\ 1\,)\qquad &\ 3\qquad&0\\
%  \hline
% \end{tabular}
%\end{center}
%\end{table}

Large degrees of freedom in the internal space for
the {EW}$\times${H} symmetry bring about necessarily complications
in dynamics of the scalar triplets $\bPhi(x)$ and $\Phi(x)$.
There exist various types of identical equations among the EW and H
invariants composed of these triplets. To construct the Lagrangian
density of the scalar triplets without redundancy, it is necessary
to examine all of the invariants and find out relations among them.
Analysis on the identical equations is made in Appendix.

\section{Lagrangian density of the V$\times$H gauge field theory}
The Lagrangian density of the theory consists of the parts being
dependent and independent of the fermion fields. In construction of the
Lagrangian density, it is tacitly understood that the scalar products are
taken, without explicitly specifying symbols for the operations, to form
the invariants out of {V} and {H} multiplets.

%\subsection{Lagrangian density depending on fermion fields}
The fermionic Lagrangian density is the sum of the kinetic
terms including gauge-interactions, $\cL_{\Psi A \bA}$, and the
interaction terms between the fermion and scalar fields, $\cL_{\rm fs}$.
%, as
%\begin{equation}
%    \cL_{\rm f} = \cL_{\Psi A \bA} + \cL_{\rm fs}\;.
%\end{equation}
The kinetic part $\cL_{\Psi A \bA}$ consists of the bilinear
form of the chiral fermion fields as follows:
\begin{equation}
   \cL_{\Psi A \bA} = 
   \sum_{f,h}\,\bar{\Psi}_h^{f}(x)\,i\gm^\mu\cD_\mu\,\Psi_h^{f}(x)
\end{equation}
where $\cD_\mu$ is the covariant derivative
\begin{equation}
  \hspace*{-0.5cm}
  \cD_\mu = \pd_\mu
          - i\{{\rm V\ gauge\ fields}\}_\mu
          - i\{{\rm H\ gauge\ fields}\}_\mu
  \label{covderivfermion}
\end{equation}
which includes the gauge fields for both of the V and H symmetries.
%The second term in (\ref{covderivfermion}) is given by
%\begin{equation}
%  \{{\rm V\ gauge\ fields}\}_\mu
%  = g_3 A_\mu^{(3)j}(x)\,\half\lambda_j
%  + g_2 A_\mu^{(2)j}(x)\,\half\tau_j
%  + g_1 A_\mu^{(1)}(x)\,\half\,Y
%\end{equation}
%where $Y$ is the EW hypercharge operator.
The third term in (\ref{covderivfermion}) is expressed in terms of the gauge
fields $\bA_\mu^{(2)j}(x)$ for $SU_{\rm H}(2)$ group and the generators
$\btau_j$ of the algebra $\bcA$ as
\begin{equation}
  \{{\rm H\ gauge\ fields}\}_\mu
  = \bg_2\,\bA_\mu^{(2)j}(x)\,\half\,\btau_j
\end{equation}
where the gauge field $\bA_\mu^{(1)}(x)$ does not exist, since the zero
H hypercharge is assigned to all of the fermionic triplets.

The Lagrangian density $\cL_{\rm fs}$ consists of the fermion-scalar
interactions of the Yukawa and Majorana types as
\begin{equation}
  \cL_{\rm fs} = \sum_{f=u,\,d,\,\nu,\,e}\cL_{\rm Y}^{\,f} + \cL_{\rm M}
\end{equation}
where the Yukawa part $\cL_{\rm Y}^{\,f}$ includes the fermion triplet of
the $f$-sector $(f=u,\,d,\,\nu,\,e)$ and the Majorana part $\cL_{\rm M}$ is
composed of the triplet $\Psi_R^{\nu}(x)$ of right-handed neutrino fields.
The $SU_{\rm H}(2)$ invariance of the quantities $\ok\Psi_h^f(x)\ck$, 
$\ok\Phi(x)\ck$ and $\ok\bPhi(x)\ck$ brings about ingenious structure
for the Yukawa and Majorana interactions. 

%\subsubsection{Yukawa interaction}
The density $\cL_{\rm Y}^f$ is constructed by summing up all of
the {EW}$\times${H} invariants consisting of bilinear forms of fermion
triplets and the scalar triplets $\Phi(x)$ and $\wtPhi(x)$. We find
\begin{equation}
 \begin{array}{ll}
 \cL_{\rm Y}^{\,f}
  &=Y_{f1}\,\bar{\Psi}_L^{\,f^{\pr}}\,\wtPhi\,\ok\Psi_R^{\,f}\ck
  +Y_{f2}\,\ok\bar{\Psi}_L^{\,f^{\pr}}\ck\,{}^{\bt}\wtPhi i\btau_2\Psi_R^{\,f}\\
 \noalign{\vskip 0.2cm}
  &+\, Y_{f3}\,\bar{\Psi}_L^{\,f^{\pr}}i\tau_2\ok\Phi^\ast\ck\Psi_R^{\,f}
   +Y_{f4}\,\ok\bar{\Psi}_L^{\,f^{\pr}}\ck
            i\tau_2\ok\Phi^\ast\ck\ok\Psi_R^{\,f}\ck
   + {\rm h.c.}
 \end{array}
  \label{YukawaUp}
\end{equation}
for the EW up-sectors $(f^\pr = q,\,f=u)$ and $(f^\pr = l,\,f=\nu)$, and
\begin{equation}
  \begin{array}{ll}
  \cL_{\rm Y}^{\,f}
  & = Y_{f1}\,\bar{\Psi}_L^{\,f^{\pr}}\,\Phi\,\ok\Psi_R^{\,f}\ck
    + Y_{f2}\,\ok\bar{\Psi}_L^{\,f^{\pr}}\ck
            \,{}^{\bt}\Phi i\btau_2 \Psi_R^{\,f}\\
   \noalign{\vskip 0.2cm}
  & +\, Y_{f3}\,\bar{\Psi}_L^{\,f^{\pr}}\ok\Phi\ck\Psi_R^{\,f}
    + Y_{f4}\,\ok\bar{\Psi}_L^{\,f^{\pr}}\ck\ok\Phi\ck \ok\Psi_R^{\,f}\ck
    + {\rm h.c.}
  \end{array}
  \label{YukawaDown}
\end{equation}
for the EW down-sectors $(f^\pr = q,\,f=d)$ and $(f^\pr = l,\,f=e)$.
Each sector includes four unknown complex coupling constants
$Y_{fi}\ (i=1,\,\cdots,\,4)$. In these densities, the operation of
the H-sum plays an indispensable role to generate $SU_{\rm H}(2)$ invariants.
Without the H-sums which reflects a unique feature of the group $\bG(\bA)$
embodied in (\ref{property3vector}), the Yukawa interaction turns out to
be empty in (\ref{YukawaUp}) and (\ref{YukawaDown}). The assignment of
zero H hypercharge for all of the triplets $\Psi^f_h(x)$ and $\Phi(x)$
is essential in the present formalism.

%In the SM, many unknown Yukawa coupling constants are main cause
%of uncertainty and complexity in the flavor physics. In the present
%scheme, the H symmetry brings forth order and constraint in
%the pattern of the Yukawa interactions and reduces considerably
%the number of the unknown parameters.

The Majorana interaction is possible only for the right-handed triplet
for neutrino species possessing no SM quantum number. To look for
H invariants, it is crucial to remark that the conservation of
H hypercharge forbid the quantity $\ok\bPhi\ck$ to appear and that
the element $\btau_2$ works properly to cancel the H hypercharge effect
of the triplet $\bPhi(x)$ owing to the expression for $\Omega_1$ in
(\ref{GroupElementinAlgebra}) and $\btau_2\bD=0$.
%
%There are two types of the Lorentz and H invariant combinations of
%the neutrino triplet $\Psi_R^{\,\nu}$. Using the symbols $t$ and $\bt$
%for transpositions with respect to the Dirac spinor and H triplet,
%respectively, we find one type of the invariant as
%\begin{equation}
%   {}^{t\bt}\Psi_R^{\,\nu}\,C\,\btau_2\Psi_R^{\,\nu}
%   \label{MajoranaInv1}
%\end{equation}
%where $C$ is the charge-conjugation matrix specified by the relations
%\begin{equation}
%   C^{-1}\,\gamma_\mu\,C = -{}^t\gamma_\mu,\quad
%   C = -C^{-1} = -C^\dag = -{}^tC\;.
%\end{equation}
%Invariance of (\ref{MajoranaInv1}) under the H symmetry transformation is
%proved by the relation (\ref{transposition}). We find also the following
%types of invariants as
%\begin{equation}
%   {}^{t\bt}\Psi_R^{\,\nu}\,C\,\btau_2\bPhi\ok\Psi_R^{\,\nu}\ck,\quad 
%   {}^t\ok\Psi_R^{\,\nu}\ck\,C\,{}^{\bt}\bPhi\btau_2\Psi_R^{\,\nu}
%\end{equation}
%
It is customary to use charge conjugates of neutrino fields to investigate
the Majorana interactions. Normally, the charge conjugates of the chiral
fermion triplets $\Psi^f_{L,R}$ are defined by
\begin{equation}
    \Psi^{fc}_{L,R} = C\,{}^{\bt t}\overline{\Psi^f_{R,L}},\ \ 
    \overline{\Psi_{L,R}^{fc}} = {}^{t\bt}\Psi_{R,L}^f\,C
\end{equation}
with the charge-conjugation matrix $C$.
The most general Lagrangian density for the Majorana interactions
consisting of EW$\times$H invariants is given in the form
\begin{equation}
\hspace*{-0.5cm}
\begin{array}{l}
\cL_{\rm M}
= {\bg}_{\rm M1}{}^{t\bt}\Psi_R^{\,\nu}\,C\,\btau_2\bPhi\ok\Psi_R^{\,\nu}\ck
 +{\bg}_{\rm M2}{}^t\ok\Psi_R^{\,\nu}\ck\,C\,
                   {}^{\bt}\bPhi\btau_2\Psi_R^{\,\nu}
 + {\bm}_{\rm M}{}^{t\bt}\Psi_R^{\,\nu}\,C\,\btau_2\,\Psi_R^{\,\nu}
 +{\rm h.c.}\\
\noalign{\vskip 0.3cm}
\quad\ \ = {\bg}_{\rm M1}
            \overline{\Psi_L^{\,\nu c}}\,\btau_2\bPhi\ok\Psi_R^{\,\nu}\ck
 +{\bg}_{\rm M2}
         \ok\overline{\Psi_L^{\,\nu c}}\ck\,{}^{\bt}\bPhi\btau_2\Psi_R^{\,\nu}
 + {\bm}_{\rm M}\overline{\Psi_L^{\,\nu c}}
        \,\btau_2\,\Psi_R^{\,\nu}+{\rm h.c.}
 \end{array}
 \label{MajoranaInt}
\end{equation}
where $\bg_{{\rm M}j} (j=1,\,2)$ and $\bm_{\rm M}$ are the Majorana coupling
constants and mass.

%\subsection{Lagrangian density independent of fermion fields}
The Lagrangian density for bosonic fields, which does not include fermion
fields, is the sum of the gauge field part and the scalar field part
$\cL_{\rm scalar}$.
%\begin{equation}
%    \cL_{\rm b} = \cL_{\rm gauge} + \cL_{\rm scalar} .
%\end{equation}
%, i.e.,
%\begin{equation}
%    \cL_{\rm gauge} = \cL^{\rm G}_{\rm V} + \cL^{\rm G}_{\rm H} .
%\end{equation}
The gauge field density is given by the Lorentz invariants of the field
strengths of the {V} and {H} gauge fields. The density of the H symmetry,
$\cL^{\rm G}_{\rm H}$, is constructed as follows:
\begin{equation}
 \cL^{\rm G}_{\rm H}
  = -\quarter\,\sum_{j=1}^3\bF^{(2)j}_{\mu\nu}\bF^{(2)j\,\mu\nu}
                 -\quarter\,\bF^{(1)}_{\mu\nu}\bF^{(1)\,\mu\nu} .
 \label{KineticHFgauge}
\end{equation}

%\subsubsection{Lagrangian density of scalar fields}
The Lagrangian density for the scalar fields $\cL_{\rm scalar}$ can be
expressed by
\begin{equation}
    \cL_{\rm scalar} = (\cD^\mu\bPhi)^\dag(\cD_\mu\bPhi)
    + (\cD^\mu\Phi)^\dag(\cD_\mu\Phi) - V_{\rm T}(\bPhi,\,\Phi)
\end{equation}
where the covariant derivatives for the scalar triplets $\bPhi(x)$ and
$\Phi(x)$ are given, respectively, by
\begin{equation}
   \cD_\mu\bPhi = \left(\pd_\mu - i\bg_2\,\bA^{(2)j}_\mu\,\half\,\btau_j
                  - i\bg_1\,\bA^{(1)}_\mu\,\by_{\small\bPhi}\bD \right)\bPhi\;,
   \label{covariantderivforbPhi}
\end{equation}
and
\begin{equation}
   \cD_\mu\Phi = \left(\pd_\mu - ig_2\,A^{(2)j}_\mu\,\half\,\tau_j
                               - ig_1\,A^{(1)}_\mu\half
                      - i\bg_2\,\bA^{(2)j}_\mu\,\half\,\btau_j \right)\Phi,
   \label{covariantderivforPhi}
\end{equation}
and $V_{\rm T}(\bPhi,\,\Phi)$ is the total Higgs potential for both of
the triplets $\bPhi(x)$ and $\Phi(x)$. Without loss of generality, we are
able to separate the potential $V_{\rm T}(\bPhi,\,\Phi)$ into three parts
as follows:
\begin{equation}
  V_{\rm T}(\bPhi,\,\Phi) = V_1(\bPhi) + V_2(\Phi) + V_3(\Phi,\,\bPhi)
\end{equation}
where $V_1(\bPhi)$ and $V_2(\Phi)$ are the potential parts of self-interactions
of the triplets $\bPhi(x)$ and $\Phi(x)$, respectively, and $V_3(\Phi,\,\bPhi)$
is the part of their mutual interactions.

To construct explicit forms of the potential parts $V_i$ without
double counting, it is necessary to use the identical relations among
the invariants of the scalar triplets investigated in Appendix.
The potential $V_1(\bPhi)$ of the self-interaction is given by
\begin{equation}
\hspace*{-0.3cm}
 \begin{array}{ll}
  V_1({\bPhi})\!\!&=
        - \bm_1^2\,\bPhi^\dag\bPhi
        - \bm_2^2\,\ok\bPhi\ck^\dag\ok\bPhi\ck
        + \half\,\brlam_1\left(\bPhi^\dag\bPhi\right)^2
        + \half\,\brlam_2\left(\ok\bPhi\ck^\dag\ok\bPhi\ck\right)^2\\
        \noalign{\vskip 0.3cm}
       &\ \ + \brlam_3\left(\bPhi^\dag\bPhi\right)
              \left(\ok\bPhi\ck^\dag\ok\bPhi\ck\right)
 \end{array}
 \label{bphipot}
\end{equation}
where $\brlam_1, \brlam_2$ and $\brlam_3$ are positive-definite coupling
constants. Using this density, we analyze the breakdown of the H symmetry,
preserving the EW symmetry, around the scale $\brLam$.
%To write down the explicit form of the potential $V_2(\Phi)$, we must
%take the conservation laws of the EW hypercharge into consideration.
Since the triplet $\Phi$ carries the EW hypercharge $Y_{\rm EW} = 1$,
the quantity $\Phi^{\dag}\wtPhi$ in (\ref{PhiTildePhi}) is not the EW
invariant. It appears in the scalar potential only as the product with
its Hermite conjugate. In consequence, we obtain
\begin{equation}
\hspace*{-0.5cm}
  \begin{array}{l}
   V_2(\Phi) =
       - m_1^2 \Phi^\dag\Phi - m_2^2 \ok\Phi\ck^\dag\ok\Phi\ck
       + \half \balam_1\left(\Phi^\dag\Phi\right)^2
       + \half \balam_2\left(\ok\Phi\ck^\dag\ok\Phi\ck\right)^2\\
  \noalign{\vskip 0.3cm}
  \qquad + \balam_3\left(\Phi^\dag\Phi\right)
                     \left(\ok\Phi\ck^\dag\ok\Phi\ck\right)
       + \balam_4|\Phi^\dag\ok\Phi\ck|^2
       + \balam_5\Phi^{\dag}i\tau_2{}^{\bt}\Phi^{\ast}{}^t\Phi i\tau_2\Phi\\
  \noalign{\vskip 0.3cm}
  \qquad +  \wtlam_1|\Phi^\dag\wtPhi|^2
         +  \wtlam_2|\wtPhi^\dag\ok\Phi\ck|^2
         +  \wtlam_3|\wtPhi^\dag(i\tau_2)\ok\Phi^\ast\ck|^2\\
  \end{array}
  \label{phipot}
\end{equation}
where, in the term with the coupling constant $\balam_5$, the H scalar
products are taken between the first $\Phi^{\dag}$ and the last $\Phi$ and
between the middle ${}^{\bt}\Phi^{\ast}$ and ${}^t\Phi$.

Finally, the potential for mutual-interactions between the triplets $\Phi(x)$
and $\bPhi(x)$ is proved to take the form:
\begin{equation}
\hspace*{-0.6cm}
  \begin{array}{l}
   V_3(\Phi,\,\bPhi) =
           \dlam_1(\Phi^\dag\Phi)(\bPhi^{\dag}\bPhi)
         + \dlam_2(\Phi^\dag\Phi)(\ok\bPhi\ck^{\dag}\ok\bPhi\ck)
         + \dlam_3(\ok\Phi\ck^\dag\ok\Phi\ck)(\bPhi^{\dag}\bPhi)\\
  \noalign{\vskip 0.3cm}
  + \dlam_4(\ok\Phi\ck^\dag\ok\Phi\ck)
                       (\ok\bPhi\ck^{\dag}\ok\bPhi\ck)
  + \dlam_5(\wtPhi^\dag\bPhi)(\bPhi^{\dag}\wtPhi)
  + \dlam_6|\Phi^\dag(I-\bD)\bPhi|^2
  + \dlam_7|\Phi^\dag\bD\bPhi|^2.
   \end{array}
   \label{bphiphipot}
\end{equation}

\section{Remark on a compound mechanism of symmetry breakdowns}
At present, there exists no established theory which can describe
consistently multi-stage-breakdowns of symmetries with vastly-different
energy scales. We formulate, here, a compound mechanism of symmetry
breakdowns that can lead, without creating unphysical modes of massless
particles and tachyons, to an effective theory with lesser numbers of
unknown parameters as possible for low energy phenomenology. For its
purpose, let us postulate first that the triplet $\bPhi(x)$ ($\Phi(x)$)
takes a stationary reference state\footnote{In this trial framework of
the symmetry breakdown, we abstain from using the concept of the
{\lq\lq}vacuum{\rq\rq} which describes the stable potential minimum of
the quantum system of the scalar multiplets.} specified by a single
parameter. And then, we decompose the triplet around the reference state
and impose necessary modifications so as to forbid unphysical modes
to appear. 

Around the high energy scale $\brLam$, the reference state of the triplet
$\bPhi$ is derived by neglecting effects of the triplet $\Phi(x)$.
Let us assume that the reference state takes the form
\begin{equation}
  \langle\, \bPhi \,\rangle
   = {}^{\bt}\left(\,0,\,0,\,\bv\,\right)
    = - \sqrt{\frac{2}{3}}\,\bv\,|\,2\,\rangle 
      + \frac{1}{\sqrt{3}}\,\bv\,|\,3\,\rangle
  \label{VEVbPhi}
\end{equation}
where $\bv \approx \brLam$. Following the standard procedure, we calculate
a stationary point of $V_1(\langle\, \bPhi \,\rangle)=V_1(\bv)$
%\begin{equation}
%  V_1(\bv) = -(\,\bm_1^2 + \bm_2^2\,)\bv^2
%  + \half\,(\brlam_1 + \brlam_2 + 2\brlam_3)\,\bv^4
%\end{equation}
by differentiating it with respect to $\bv$ and obtain
%the symmetry breaking
\begin{equation}
 \bv^2= \frac{\,\bm_1^2+\bm_2^2}{\brlam_1 + \brlam_2 + 2\brlam_3}.
 \label{VEVbv}
\end{equation}

It is natural to consider that the breakdown of the H symmetry might give
influence on the symmetry breakdown in low energy region around the scale
$\Lambda$. We look for a reference state for the triplet $\Phi(x)$
by examining a local symmetry-violating stationary-point of the potential
$V_2(\Phi)+V_3(\Phi, \bv)$. To deduce a compact effective theory for low energy,
let us assume, here again, the scalar triplet  $\Phi(x)$ takes the reference
state in the simplest form
\begin{equation}
 \langle\, \Phi \,\rangle
    = {}^{\bt}\left(\!
              \begin{array}{ccc}
                \!\left(\!
                \begin{array}{c}
                  0\\
                \noalign{\vskip 0.2cm}
                  0
                \end{array}
                \!\right),\!&\!
                \!\left(\!
                \begin{array}{c}
                  0\\
                \noalign{\vskip 0.2cm}
                  0
                \end{array}
                \!\right),\!&\!
                \!\left(\!
                \begin{array}{c}
                  0\\
                \noalign{\vskip 0.2cm}
                  v_0
                \end{array}
                \!\right)
              \end{array}
            \!\!\right)
      = -\sqrt{\frac{2}{3}}
           \left(\!
             \begin{array}{c}
               0\\
             \noalign{\vskip 0.2cm}
               v_0
             \end{array}\!
           \right)|\,2\,\rangle
        +\frac{1}{\sqrt{3}}
           \left(\!
             \begin{array}{c}
               0\\
             \noalign{\vskip 0.2cm}
               v_0
             \end{array}\!
           \right)|\,3\,\rangle
       \label{VEVPhi}
\end{equation}
where $v_0$ is a parameter specifying a local symmetry-violating
stationary-point. Taking derivative of
$V_2(\langle\, \Phi \,\rangle)+V_3(\langle\, \Phi \,\rangle,\,\bv)$
with respect to $v_0$, we obtain
%\begin{equation}
%V_2(v) = -(\,m_1^2 + m_2^2\,)v^2
% + \half\,(\balam_1+\balam_2+2\balam_3+2\balam_4
%       +\frac{4}{3}\balam_5+\frac{4}{3}\wtlam_3)\,v^4
%\end{equation}
\begin{equation}
 v_0^2= \frac{\,m_1^2+m_2^2-(\dlam_1+\dlam_2+\dlam_3+\dlam_4
       +\frac{4}{9}\dlam_6+\frac{1}{9}\dlam_7)\bv^2}
       {\balam_1+\balam_2+2\balam_3+2\balam_4+\frac{4}{3}\wtlam_3}.
 \label{v0square}
\end{equation}

Evidently, the simplest choices of the reference states in (\ref{VEVbPhi})
and (\ref{VEVPhi}) break only partially the H symmetry.
As shown explicitly in the following sections, the remaining symmetry
causes massless Nambu-Goldstone (NG) particles to appear. Fortunately,
there is a mechanism to circumvent such difficulties by utilizing the
specific property of the vector of the reference state that can be
expressed uniquely in terms of linear combination of the eigenvectors
$|2\rangle$ and $|3\rangle$, as shown in the right-hand sides of
(\ref{VEVbPhi}) and (\ref{VEVPhi}). This feature enables us to decompose
the triplet around the reference state so that one component field can be
shared in both coefficients of $|2\rangle$ and $|3\rangle$ by accompanying
an adjustable mixing parameter. Such decompositions of the triplets $\bPhi(x)$
and $\Phi(x)$ are, respectively, given in (\ref{bPhidecomposition}) and
(\ref{Phidecomposition}) below. It is this facility of sharing the component
field with mixing parameter that acts to break the residual symmetry
and suppress the appearance of massless scalar fields in the phases of
broken symmetries.

Substitution of the decompositions of the scalar triplets in the broken phase
into the Lagrangian density produces, necessarily, anomalous unphysical terms
depending linearly on some of the component scalar fields. It is possible
to set the coefficients of such harmful terms to disappear by adjusting
finite values appropriately for the mixing parameters in the decompositions
in (\ref{bPhidecomposition}) and (\ref{lowmixings}). Consequently, this
mechanism which affords non-vanishing mixing parameters breaks the residual
symmetry and enables the would-be NG modes to acquire finite masses.
%The new mixing mechanism can be applied effectively to
%the symmetry breakdowns at both the high and low energy scales.

In the low energy symmetry breakdown, however, this mixing mechanism of
component fields is not enough to forbid the decomposition of the triplet
$\Phi(x)$ around the reference state in (\ref{Phidecomposition}) to have
a component field with an imaginary mass. Another device is necessary to
complete the compound mechanism of the symmetry breakdowns. To suppress
a tachyonic mode to appear, we have to utilize a freedom of rescaling of
the parameter $v_0$ as made in (\ref{rescaling}).

\section{Symmetry breakdown at high energy scale $\brLam$}
%To examine the breakdown of the H symmetry, we must recognize that
%the EW symmetry holds around the high energy scale $\brLam$. Therefore,
%it is reasonable to apply the generalized Higgs mechanism to the potential
%$V_1(\bPhi)$ in (\ref{bphipot}).
%
%\subsection{Gauge-fixing of the H symmetry}
In the broken phase of the H symmetry around the scale $\brLam = \bv$,
the scalar triplet field $\bPhi(x)$ can be decomposed into
\begin{equation}
  \bPhi(x) = \Omega(\bvth(x))\,\bPhi_0(x)
  \label{HFUnitaryPhaseofbPhi}
\end{equation}
where $\Omega(\bvth(x))$ is the unitary group element of the H symmetry
including the local fields $\bvth_j(x)\,(j=0,\,1,\,2,\,3)$ destined to be
gauged away. The gauge-fixed part $\Phi_0(x)$ is assumed to take
the following form as
\begin{equation}
  \bPhi_0(x)
   = \frac{1}{\sqrt{2}}\bxi_1(x)|\,1\,\rangle
    + \left(\frac{1}{\sqrt{2}}\balph\,\bxi_2(x)
    - \sqrt{\frac{2}{3}}\bv\right)|\,2\,\rangle
    + \left(\frac{1}{\sqrt{2}}\bbeta\,\bxi_2(x)
    + \frac{1}{\sqrt{3}}\bv\right)|\,3\,\rangle
  \label{bPhidecomposition}
\end{equation}
in which $\bxi_1(x)$ and $\bxi_2(x)$ are real scalar fields, and
$\balph=\cos\bth$ and $\bbeta=\sin\bth$ are parameters for mixing.
With this gauge-fixing of the scalar triplet $\bPhi(x)$, the gauge
fields are metamorphosed to the massive vector fields through
the local gauge transformation
$\bA_\mu(x)\rightarrow \bA_\mu^\pr(x)$ defined by
\begin{equation}
   \cD_\mu(\bA^\pr)=\Omega^{-1}(\bvth(x)) \cD_\mu(\bA) \Omega(\bvth(x))
       \label{HFUnitaryPhaseofbA}
\end{equation}
where $\cD_\mu(\bA)$ is the covariant derivative for the scalar triplet
$\bPhi(x)$ in (\ref{covariantderivforbPhi}).

The gauge-fixing of the H symmetry imposed on the scalar triplet $\bPhi(x)$
gives necessarily influences on all other fields.
%The mutual interaction between the scalar triplets with coupling constants
%$\dlam_j$ in the potential (\ref{bphiphipot}) requires the following
%factorization of triplet $\Phi(x)$ as
%\begin{equation}
%       \Phi(x) = \Omega(\bvth(x))\uPhi(x)
%       \label{HFUnitaryPhaseofPhi}
%\end{equation}
%where $\Omega(\bvth(x))$ is the same unitary factor in the decomposition
%(\ref{HFUnitaryPhaseofbPhi}). 
%the unitary phase factor $\Omega(\bvth(x))$ 
%of H symmetry must also be separated from all of the fermion triplets at
%the scale $\brLam$ as
%\begin{equation}
%       \Psi_h^f(x) = \Omega(\bvth(x))\uPsi_h^f(x).
%       \label{HFUnitaryPhaseofPsi}
%\end{equation}
Namely, it necessitates inevitably to adjust the same unitary factor
$\Omega(\bvth(x))$ for the scalar and fermion triplets, $\Phi(x)$
and $\Psi_h^f(x)$, leaving the gauge-fixed parts $\uPhi(x)$ and
$\uPsi_h^f(x)$ as follows:
\begin{equation}
       \Phi(x) = \Omega(\bvth(x))\uPhi(x),\quad
       \Psi_h^f(x) = \Omega(\bvth(x))\uPsi_h^f(x).
       \label{HFUnitaryPhaseofPhiPsi}
\end{equation}

%\subsection{Scalar fields descended from $\bPhi(x)$}
Substitution of the decomposition (\ref{HFUnitaryPhaseofbPhi}) and
(\ref{bPhidecomposition}) into the potential
$V_1(\bPhi)$ in (\ref{bphipot}) leads to
\begin{equation}
\hspace*{-0.2cm}
 \begin{array}{l}
  V_1(\bPhi_0)\!=\!
      \bv\,\left\{\sqrt{\frac{2}{3}}
      \left[\bm_1^2-(\brlam_1+\brlam_3)\bv^2
      \right](\sqrt{2}\balph-\bbeta)
     -\sqrt{6}\left[\bm_2^2-(\brlam_2+\brlam_3)\bv^2\right]
      \bbeta\right\}\,\bxi_2\\
  \noalign{\vskip 0.2cm}
  \quad+\ \frac{1}{2}[-\bm_1^2+(\brlam_1+\brlam_3)\bv^2]
                \left(\bxi_1^2+\bxi_2^2\right)\\
  \noalign{\vskip 0.2cm}
  \quad+\ \half\left\{-3\bbeta^2\bm_2^2
      + \left[\frac{2}{3}\brlam_1(\sqrt{2}\balph-\bbeta)^2
             +3(3\brlam_2+\brlam_3)\bbeta^2
             -4\brlam_3(\sqrt{2}\balph-\bbeta)\bbeta\right]\bv^2
      \right\}\bxi_2^2\\
  \noalign{\vskip 0.2cm}
   \quad-\ \sqrt{\frac{2}{3}}\left[\brlam_1(\sqrt{2}\balph-\bbeta)
       -\frac{3}{2}\brlam_3\bbeta\right]\bv
       \left(\bxi_1^2+\bxi_2^2\right)\bxi_2
      + \sqrt{\frac{3}{2}}\left[3\brlam_2\bbeta
            - \brlam_3(\sqrt{2}\balph-\bbeta)\right]\bbeta^2\bv\bxi_2^3 \\
  \noalign{\vskip 0.2cm}
   \quad+\  \frac{1}{8}\brlam_1\left(\bxi_1^2+\bxi_2^2\right)^2 
      + \frac{9}{8}\brlam_2\bbeta^4\bxi_2^4 
      + \frac{3}{4}\brlam_3\bbeta^2\left(\bxi_1^2+\bxi_2^2\right)\bxi_2^2
      + V_1(\bv) .
 \end{array}
\end{equation}
The first term of this reduced potential is linear with respect to
the field $\bxi_2(x)$. Postulating the coefficient of this unphysical
term to vanish, we obtain the condition which fixes the parameter $\bth$
%\begin{equation}
%      \left[\bm_1^2-(\brlam_1+\brlam_3)\bv^2
%      \right](\sqrt{2}\balph-\bbeta)
%     -3\left[\bm_2^2-(\brlam_2+\brlam_3)\bv^2\right]\bbeta = 0
%     \label{constrainthigh}
%\end{equation}
as follows:
\begin{equation}
     \tan\bth = \sqrt{2}\frac{\bm_1^2-(\brlam_1+\brlam_3)\bv^2}
               {\bm_1^2+3\bm_2^2-(\brlam_1+3\brlam_2+4\brlam_3)\bv^2} .
     \label{tanbtheta}
\end{equation}

The masses of the real scalar fields $\bxi_1(x)$ and $\bxi_2(x)$ are
calculated to be
\begin{equation}
   m_{\tbxi_1}^2 = - \bm_1^2 + (\brlam_1+\brlam_3)\bv^2 \propto \sin\bth
  \label{massbxi1}
\end{equation}
and
\begin{equation}
 \begin{array}{l}
   m_{\tbxi_2}^2 = m_{\tbxi_1}^2 - 3\bm_2^2\sin^2\bth
    + \disp\frac{1}{3}
            \left[\,4\brlam_1\cos^2\bth-2\sqrt{2}(\brlam_1+3\brlam_3)\sin2\bth
            \right.\\
   \noalign{\vskip 0.2cm}\hspace*{5cm}
   \left.+\,(2\brlam_1+27\brlam_2+21\brlam_3)\sin^2\bth\,\right]\bv^2.
 \end{array}
  \label{massbxi2}
\end{equation}
The mass of the field $\bxi_1(x)$ is proportional to $\sin\bth$. If $\bth=0$,
the field $\bxi_1(x)$ remain necessarily in a massless mode. Therefore,
the sharing of the component field $\bxi_2(x)$ in both of the coefficients
of $|\,2\,\rangle$ and $|\,3\,\rangle$ in (\ref{bPhidecomposition}) is
indispensable to break the residual symmetry and set free the fields
$\bxi_1(x)$ from the NG theorem.

%\subsection{Vector fields originated from the H gauge fields}
To derive configurations of massive vector fields $\bA_\mu^\pr(x)$ which
are related with the gauge fields $\bA_\mu(x)$ by the transformation in
(\ref{HFUnitaryPhaseofbA}), it is sufficient to calculate the action of
the covariant derivative on the reference state in (\ref{VEVbPhi})
as follows:
\begin{equation}
 \begin{array}{ll}
  \cD_\mu(\bA^\pr) \langle\, \bPhi\,\rangle \!\!\!
   &= \left(\pd_\mu - i\bg_2\bA^{(2){\pr}j}_\mu\,\half\,\btau_j
    - i\bg_1\,\bA^{(1){\pr}}_\mu\,\by_{\bPhi}\bD \right)
      \langle\,\bPhi\,\rangle \\
      \noalign{\vskip 0.3cm}
   &=  i \frac{1}{2}
        \left(
         \begin{array}{r}
          \sqrt{2}M_{\bW}\bW_\mu - M_{\bY}\bY_\mu \\
          \noalign{\vskip 0.2cm}
          -\sqrt{2}M_{\bW}\bW_\mu - M_{\bY}\bY_\mu \\
          \noalign{\vskip 0.2cm}
           \sqrt{2}M_{\bZ}\bZ_\mu \\
          \end{array}\ 
        \right) \\
 \end{array}
 \label{bAtobWbYbZ}
\end{equation}
where the vector fields $\bW_\mu(x)$, $\bY_\mu(x)$ and $\bZ_\mu(x)$ are
expressed, tentatively, by
\begin{equation}
\disp \bW_\mu = \frac{\bA^{(2){\pr}1}_\mu - i\bA^{(2){\pr}2}_\mu}{\sqrt{2}}, \ 
      \bY_\mu = \frac{\bg_2\bA^{(2){\pr}3}_\mu
                + 2\bg_1\by_{\small\bPhi}\bA^{(1){\pr}}_\mu}
                          {\sqrt{\bg_2^2+4\bg_1^2\by_{\small\bPhi}^2}}, \ 
      \bZ_\mu = \frac{\bg_2\bA^{(2){\pr}3}_\mu
                - \bg_1\by_{\small\bPhi}\bA^{(1){\pr}}_\mu}
                          {\sqrt{\bg_2^2+\bg_1^2\by_{\small\bPhi}^2}},
      \label{WYZconfigutations}
\end{equation}
and their masses are given, respectively, by
$M_{\bW}^2 = \frac{1}{3}\bg^2\bv^2$,\,
$M_{\bY}^2 = \frac{1}{9}\left(\bg_2^2+4\bg_1^2\by_{\small\bPhi}^2\right)\bv^2$
and $M_{\bZ}^2=\frac{2}{9}\left(\bg_2^2+\bg_1^2\by_{\small\bPhi}^2\right)\bv^2$.

So far, no restriction is assumed to exist among the gauge coupling constants
$\bg_2$ and $\bg_1$ and the H hypercharge $\by_{\small\bPhi}$. Consistency of
the theory, however, requires an additional relation among them. Namely,
calculation of kinetic terms of the vector fields $\bY_\mu(x)$ and
$\bZ_\mu(x)$ by substituting the inverse relations of
(\ref{WYZconfigutations}) into the Lagrangian density
$\cL_{\rm H}^{\rm G}(\bA_\mu)=\cL_{\rm H}^{\rm G}(\bA_\mu^{\pr})$
in (\ref{KineticHFgauge}) proves that, there arise unphysical kinetic terms
such as $\pd_\mu\bY_\nu\,\pd^\mu\bZ^\nu$, unless the relation
\begin{equation}
    \bg_2^2 = 2 \bg_1^2 \by^2_{\small\bPhi}
    \label{Conditionforby}
\end{equation}
holds. This condition allows to relate the fields $\bY_\mu(x)$ and
$\bZ_\mu(x)$ with the fields $\bA_\mu^{(2)\pr 3}(x)$ and
$\bA_\mu^{(1)\pr}(x)$ by the orthogonal transformation
\begin{equation}
\disp \bY_\mu = \frac{1}{\sqrt{3}}\bA^{(2){\pr}3}_\mu
                + \sqrt{\frac{2}{3}}\bA^{(1){\pr}}_\mu,\ \ 
      \bZ_\mu = \sqrt{\frac{2}{3}}\bA^{(2){\pr}3}_\mu
                - \frac{1}{\sqrt{3}}\bA^{(1){\pr}}_\mu
\end{equation}
and results in the degenerate masses, i.e.,
$M_{\bY}^2 = M_{\bZ}^2 = M_{\bW}^2 = \frac{1}{3}\bg_2^2\bv^2$.

%Then, substituting the decomposition of the scalar triplet $\bPhi(x)$
%in (\ref{bPhidecomposition}) into the kinetic part of the Lagrangian density
%$\cL(\bPhi)$, we find the quadratic parts of the new scalar and vector
%fields in the form
%\begin{equation}
% \begin{array}{lcl}
%  (\cD_\mu\bPhi)^\dag(\cD^\mu\bPhi) &=& \half\pd_\mu\bxi_1\pd^\mu\bxi_1
%   +  \half\pd_\mu\bxi_2\pd^\mu\bxi_2 \\
%   \noalign{\vskip 0.3cm}
%   &&+ M_{\bW}^2\,\bW_\mu\bW^\mu
%   + \half\,M_{\bY}^2\,\bY_\mu\bY^\mu + \half\,M_{\bZ}^2\,\bZ_\mu\bZ^\mu
%   + \cdots
% \end{array}
%\end{equation}
%where the ellipsis signifies interactions among the new massive vector and
%scalar fields.

In this way, the compound mechanism of symmetry breakdown at the scale
$\brLam$ succeeds to transform all of the massless gauge and scalar fields
into the massive vector and scalar fields. To accomplish these results, it
is necessary to assume that the triplet $\bPhi(x)$ possesses a non-vanishing
H hypercharge satisfying the relation in (\ref{Conditionforby}), and that
the mixing angle of its decomposition in (\ref{bPhidecomposition}) takes
the definite value obeying (\ref{tanbtheta}).

Through the symmetry-breaking at the scale $\brLam$, the freedoms of
the triplet $\bPhi(x)$ are transferred to the gauge fields. A balance sheet
of transferring of the field degrees of freedom at this phase transition
is schematically summarized as follows:
\begin{equation}
\begin{array}{c}
  \left\{\,\ 
   \begin{array}{cl}
    4 {\rm \ gauge\ fields\ } \bA^{(2)i}_\mu(x) \,(i=1,\,2,\,3)
      {\rm \ and\ } \bA^{(1)}_\mu(x) & :\ 4\times 2\\
     \noalign{\vskip 0.1cm}
    3 {\rm \ massless\ complex\ scalar\ fields\ }
     \bphi_{i}(x)\, (i=1,\,2,\,3) & :\ 3\times 2
   \end{array}\ 
  \right\} \\
  \noalign{\vskip 0.1cm}
  \Downarrow \\
  \noalign{\vskip 0.1cm}
  \left\{\,\ 
   \begin{array}{cl}
     1 {\rm \ massive\ complex\ vector\ fields\ } \bW_\mu(x)
     & :\ 2\times 3 \\
     \noalign{\vskip 0.1cm}
     2 {\rm \ massive\ real\ vector\ fields\ }
     \bY_\mu(x){\rm \ and \ }\bZ_\mu(x)
     & :\ 2\times 3 \\
     \noalign{\vskip 0.1cm}
     2 {\rm \ massive\ real\ scalar\ fields\ } \bxi_{i}(x)\,(i=1,\,2)
     & :\ 2\times 1 
   \end{array}\,\ 
  \right\}.
\end{array}
\label{BalancebPhi}
\end{equation}
Here the number of modes of independent fields is preserved as $4 \times 2 +
3 \times 2 = 14 = 2 \times 3 + 2 \times 3 + 2\times 1$ before and after
the phase transition.

%\subsection{Majorana mass matrix for neutrino species}
Substitution of the decomposition of $\bPhi_0$ in (\ref{bPhidecomposition})
into (\ref{MajoranaInt}) leads to the effective Lagrangian density of neutrino
species as
\begin{equation}
 \cL_{\rm M}\ \rightarrow\ 
  \cL^{\rm M}_{\cM} = \overline{\uPsi^{\nu c}_L}\breve{\cM}_\nu\uPsi^\nu_R
                    + {\rm h.c.} + \cdots
\end{equation}
where the ellipsis stands for the interactions of neutrinos with the scalar
fields $\bxi_i(x)$, and $\breve{\cM}_{\nu}$ is the Majorana mass matrix
\begin{equation}
   \breve{\cM}_{\nu} = 
               \frac{1}{\sqrt{3}}B_{\nu 1}
                     \left(
                      \begin{array}{rrr}
                         1 &  1 &  1  \\
                        -1 & -1 & -1  \\
                         0 &  0 &  0  \\
                      \end{array}
                     \right)
             + \frac{1}{\sqrt{3}}B_{\nu 2}
                      \left(
                       \begin{array}{ccc}
                         1 & -1 & 0 \\
                         1 & -1 & 0 \\
                         1 & -1 & 0 \\
                       \end{array}
                     \right)
             + C_{\nu}\,\btau_2
   \label{Majoranamass}
\end{equation}
in which the coefficients are given by $B_{\nu 1} = \bg_{\rm M1}\bv$,
$B_{\nu 2} = \bg_{\rm M2}\bv$ and $C_{\nu} = \bm_{\rm M}$. Note that
this mass matrix is characterized by the same number of parameters with
the coupling constants in the Majorana interaction.

\section{Symmetry breakdown at low energy scale $\Lambda$}
In the energy region below and close the scale $\brLam$, it should be
the reduced Lagrangian density that describes dynamics of the triplet
$\uPhi(x)$ which loses the unitary phase factor of the H symmetry
in (\ref{HFUnitaryPhaseofPhiPsi}). Subsequently, to go down to the low
energy region around the scale $\Lambda$, effects of the renormalization
group must be taken into account for all of the physical quantities.
In particular, all coupling constants run down to the scale $\Lambda$.
For the sake of simplicity, the same symbols are used here for the
quantities including all these effects. To make analysis on the EW symmetry
breakdown around the scale $\Lambda$, we have to examine the potential
$V_2(\uPhi)$ and the part
\begin{equation}
\hspace*{-0.5cm}
  \begin{array}{l}
    V_3(\uPhi, \bv) = (\dlam_1+\dlam_2)\bv^2\uPhi^\dag\uPhi
                    + (\dlam_3+\dlam_4)\bv^2\ok\uPhi\ck^\dag\ok\uPhi\ck\\
    \noalign{\vskip 0.3cm}
    \quad\ \ 
     + \frac{2}{3}\dlam_5\bv^2|\langle\underline{\Phi}|1\rangle|^2
     + \frac{2}{3}\dlam_6\bv^2|\langle\underline{\Phi}|2\rangle|^2
     + \frac{1}{3}\dlam_7\bv^2|\langle\underline{\Phi}|3\rangle|^2
  \end{array}
   \label{V3LowEnergyHiggs}
\end{equation}
which reflects the influence of the H symmetry breakdown.

%The ellipsis in (\ref{V3LowEnergyHiggs}) represents residual
%terms of interactions between the component fields of $\Phi(x)$ and
%the real scalar fields $\bxi_1(x)$ and $\bxi_2(x)$. The field $\bxi_1(x)$
%which does not interact fermions except for neutrinos can couple with
%other bosonic fields produced through the symmetry breakdowns.

%\subsection{Scalar fields descended from the triplet $\Phi(x)$}
To investigate behaviors of the triplets $\uPhi(x)$ and $\uPsi(x)$ in
the broken phase of the EW symmetry, it is necessary further to separate
the unitary gauge factor as
\begin{equation}
    \uPhi(x) = \Omega_{\rm EW}(\vartheta(x))\Phi_0(x),\ \ 
    \uPsi_h^f(x) = \Omega_{\rm EW}(\vartheta(x))\Psi_{h0}^f(x)
    \label{EWUnitaryPhase}
\end{equation}
where $\Omega_{\rm EW}(\vartheta(x))$ is a group element of the EW symmetry,
which includes local component fields to be gauged away. As noticed in \S 5,
the decomposition of the triplet $\Phi_0$ around the state in (\ref{VEVPhi})
has a danger to bring about component fields of zero and imaginary masses.
In order to suppress the tachyonic mode to appear, we rescale the parameter 
$v_0$ into a new value by
\begin{equation}
          v^2 = Z\,v_0^2\ \ (\,Z>1\,)
          \label{rescaling}
\end{equation}
and introduce the following decomposition of the $\Phi_0(x)$ as
\begin{equation}
          \Phi_0(x) = \left(
             \begin{array}{c}
              \zeta_1^{+}(x)\\
              \noalign{\vskip 0.2cm}
              \zeta_1^{0}(x)
             \end{array}
           \right)|\,1\,\rangle
         + \left(
             \begin{array}{c}
              \zeta_2^{+}(x)\\
              \noalign{\vskip 0.2cm}
              \zeta_2^{0}(x)-\sqrt{\frac{2}{3}}v
             \end{array}
           \right)|\,2\,\rangle
         + \left(
             \begin{array}{c}
               0 \\
              \noalign{\vskip 0.2cm}
              \zeta_3^0(x)+\frac{1}{\sqrt{3}}v
             \end{array}
            \right)|\,3\,\rangle
    \label{Phidecomposition}
\end{equation}
where $\zeta^{+}_i(x)\,(i=1,\,2)$ and $\zeta^{0}_i(x)\ (i=1,\,2,\,3)$ are
complex scalar component fields. As confirmed below, the condition $Z>1$
is required to exclude a tachyonic mode.

%Accordingly, the associated triplet $\wtPhi(x)$ and the H-sum $\ok\Phi(x)\ck$
%have the decompositions as follows:
%\begin{equation}
% \wtPhi(x) = \left(
%             \begin{array}{c}
%              \zeta_2^{0\ast}(x) - \sqrt{\frac{2}{3}}\,v\\
%              \noalign{\vskip 0.2cm}
%              -\zeta_2^{-}(x)
%              \end{array}
%             \right)|\,1\,\rangle
%           - \left(
%             \begin{array}{c}
%              \zeta_1^{0\ast}(x)\\
%              \noalign{\vskip 0.2cm}
%              -\zeta_1^{-}(x)
%             \end{array}
%             \right)|\,2\,\rangle
%         \label{wtPhidecomposition}
%\end{equation}
%and
%\begin{equation}
% \ok\Phi(x)\ck =
%    \left(
%      \begin{array}{c}
%             0 \\
%      \noalign{\vskip 0.2cm}
%        \sqrt{3}\,\zeta_3^0(x)+v
%      \end{array}
%    \right) .
% \label{HFSumofPhi}
%\end{equation}

The representation of $\Phi(x)$ in (\ref{HFUnitaryPhaseofPhiPsi})
(\,(\ref{EWUnitaryPhase}) and (\ref{Phidecomposition})\,) must be substituted
into the sum of the potentials $V_2(\Phi)$ in (\ref{phipot})
and $V_3(\Phi, \bv)$ in (\ref{V3LowEnergyHiggs}). Up to the second order
with respect to the component scalar fields, we obtain
\begin{equation}
 \begin{array}{l}
  V_2(\zeta)+V_3(\zeta, \bv) =
         -\sqrt{\frac{2}{3}}\,v\,\underline{[-m_1^{\pr 2}
         +(\balam_1+\balam_3+\balam_4+\wtlam_3)v^2
         +\frac{2}{3}\dlam_6\bv^2](\zeta_2^{0\ast}+\zeta_2^{0})}\\
   \noalign{\vskip 0.3cm}
   \quad+\sqrt{\frac{1}{3}}\,v\,\underline{[-m_1^{\pr 2}-3m_2^{\pr 2}
         +(\balam_1+3\balam_2+4\balam_3
           +4\balam_4+2\wtlam_3)v^2+\frac{1}{3}\dlam_7\bv^2]
          (\zeta_3^{0\ast}+\zeta_3^{0})}\\
   \noalign{\vskip 0.3cm}
   \quad+\ [-m_1^{\pr 2}+(\balam_1+\balam_3+\wtlam_2)v^2]
          (|\zeta_1^+|^2 + |\zeta_2^+|^2)
          -2\balam_5v_0^2(|\zeta_1^+|^2+\frac{1}{3}|\zeta_2^+|^2)\\
   \noalign{\vskip 0.3cm}
   \quad+\ \frac{8}{3}\wtlam_1v^2|\zeta^+_1|^2
        + [-m_1^{\pr 2}+(\balam_1+\balam_3+\balam_4+\wtlam_3)v^2]
          (|\zeta_1^0|^2 +|\zeta_2^0|^2)\\
   \noalign{\vskip 0.3cm}
   \quad+\ \frac{2}{3}\dlam_5\bv^2(|\zeta_1^+|^2 + |\zeta_1^0|^2)
        +\ \frac{2}{3}\dlam_6\bv^2(|\zeta_2^+|^2 + |\zeta_2^0|^2)
        +\ \frac{1}{3}\dlam_7\bv^2|\zeta_3^0|^2\\
   \noalign{\vskip 0.3cm}
   \quad+\ \left[-m_1^{\pr 2}-3m_2^{\pr 2}
   +(\balam_1+3\balam_2+4\balam_3+4\balam_4
                         +2\wtlam_3)v^2)\right]|\zeta_3^0|^2\\
 \noalign{\vskip 0.3cm}
   \quad+\ \frac{1}{3}\balam_1v^2(\zeta_2^{0\ast}+\zeta_2^{0})^2
   +\left(\frac{1}{6}\balam_1+\frac{3}{2}\balam_2
         +\balam_3+\balam_4\right)v^2
   (\zeta_3^{0\ast}+\zeta_3^{0})^2\\
 \noalign{\vskip 0.3cm}
   \quad-\sqrt{2}(\frac{1}{3}\balam_1+\balam_3+\balam_4+\wtlam_3)v^2
    (\zeta_2^{0\ast}+\zeta_2^{0})(\zeta_3^{0\ast}+\zeta_3^{0})
   + \cdots
   \end{array} 
   \label{LowEnergyHiggs}
\end{equation}
where 
\begin{equation}
    m_1^{\pr 2}=m_1^{2}-(\brlam_1+\brlam_2)\bv^2,\quad
    m_2^{\pr 2}=m_2^{2}-(\brlam_3+\brlam_4)\bv^2.
    \label{effectiveHiggsmass}
\end{equation}
The underlined parts are unphysical harmful terms depending linearly on
$\zeta^{0\ast}_2(x)+\zeta^{0}_2(x)$ and $\zeta^{0\ast}_3(x)+\zeta^{0}_3(x)$.
To eliminate such terms, let us reexpress the component fields $\zeta^{0}_2(x)$
and $\zeta^{0}_3(x)$ by
\begin{equation}
   \zeta_2^0(x) = \frac{1}{\sqrt{2}}[\,\eta_1(x)\cos\th + i\eta_2(x)],\quad
   \zeta_3^0(x) = \frac{1}{\sqrt{2}}[\,\eta_1(x)\sin\th + i\eta_3(x)]
   \label{lowmixings}
\end{equation}
in terms of new real fields $\eta_i(x)\,(i=1,\,2,\,3)$ and a mixing angle
$\th$. Then the underlined terms in (\ref{LowEnergyHiggs}) are absorbed
into a single term depending linearly on the new field $\eta_1(x)$.
Requiring such a harmful term to vanish, we find the constraint
which fixes the mixing angle $\th$ as
\begin{equation}
 \tan\theta = \sqrt{2}\frac{m_1^{\pr 2}
     -(\balam_1+\balam_3+\balam_4+\wtlam_3)v^2+\frac{2}{3}\dlam_6\bv^2}
     {m_1^{\pr 2}+3m_2^{\pr 2}
     -(\balam_1+3\balam_2+4\balam_3
             +4\balam_4+2\wtlam_3)v^2+\frac{1}{3}\dlam_7\bv^2} .
 \label{ratiotheta}
\end{equation}

Consequently, the second stage breakdown of the symmetry induces the three
complex fields $\zeta^{+}_1(x)$, $\zeta^{+}_2(x)$ and $\zeta^0_1(x)$, and
the three real fields $\eta_i(x)\, (i=1,\,2,\,3)$. 
The complex fields possess the masses
\begin{equation}
\begin{array}{l}
 m_{\zeta_1^{+}}^2=
 -m_1^{\pr 2}+(\balam_1+\balam_3-2\balam_5
             +\frac{8}{3}\wtlam_1+\wtlam_2)v^2+\frac{2}{3}\dlam_5\bv^2, \\
 \noalign{\vskip 0.2cm}
 m_{\zeta_2^{+}}^2=-m_1^{\pr 2}
 +(\balam_1+\balam_3-\frac{2}{3}\balam_5+\wtlam_2)v^2
 +\frac{2}{3}\dlam_6\bv^2, \\
 \noalign{\vskip 0.2cm}
 m_{\zeta_1^0}^2=
 -m_1^{\pr 2}+(\balam_1+\balam_3+\balam_4+\wtlam_3)v^2
             +\frac{2}{3}\dlam_5\bv^2 .
 \end{array}
  \label{massscalarlow1}
\end{equation}
The masses of the three real fields are calculated to be
\begin{equation}
 \begin{array}{l}
 m_{\eta_1}^2 = m_{\eta_2}^2\cos^2\theta + m_{\eta_3}^2\sin^2\th \\
  \noalign{\vskip 0.2cm}
    \qquad\ \ +\,4\left[\frac{1}{3}\balam_1\cos^2\th
    +\left(\frac{1}{6}\balam_1+\frac{3}{2}\balam_2
           +\balam_3+\balam_4\right)\sin^2\th\right. \\
  \noalign{\vskip 0.2cm}
    \qquad\ 
    \left.\qquad -\sqrt{2}\left(\frac{1}{3}\balam_1+\balam_3
           +\balam_4+\wtlam_3\right)\cos\th\sin\th\right]v^2\\
  \noalign{\vskip 0.2cm}
    \qquad\ \ +\,(\frac{2}{3}\dlam_6\cos^2\theta
                 +\frac{1}{3}\dlam_7\sin^2\theta)\bv^2,\\
  \noalign{\vskip 0.2cm}
 m_{\eta_2}^2 = -m_1^{\pr 2} +(\balam_1+\balam_3+\balam_4
     +\wtlam_3)v^2+\frac{2}{3}\dlam_6\bv^2 \, \propto\, \sin\theta,\\
  \noalign{\vskip 0.2cm}
  m_{\eta_3}^2 = -m_1^{\pr 2} - 3m_2^{\pr 2}
  +(\balam_1+3\balam_2+4\balam_3+4\balam_4
     + 2\wtlam_3)v^2 +\frac{1}{3}\dlam_7\bv^2.
 \end{array}
 \label{massscalarlow2}
\end{equation}
Note that $m^2_{\eta_2}$ is proportional to $\sin\th$. This means that
the real scalar field $\eta_2(x)$ corresponds to the would-be NG boson,
since it becomes massless if $\th = 0$.

These expressions for the squared masses of six boson fields include
more than six unknown adjustable coupling constants. Therefore,
it seems possible to make all of the squared masses to be positive by
choosing properly the values of the coupling constants. This is, however,
not the case. The expression for $v_0^2$ in (\ref{v0square}) results readily
in the identity
\begin{equation}
  2m_{\eta_2}^2+m_{\eta_3}^2 = (Z-1)(3\balam_1+3\balam_2+6\balam_3
                           +6\balam_4+4\wtlam_3)v^2
  \label{Zcondition}
\end{equation}
between $m_{\eta_2}^2$ and $m_{\eta_3}^2$. Therefore, if $Z=1$, either
$m_{\eta_2}^2$ or $m_{\eta_3}^2$ must be negative. To exclude such a tachyonic
mode, it is necessary to rescale the value $v_0$ to $v$ and impose
the condition $Z>1$.

%\subsection{Vector fields originated from the EW gauge fields}
Let us indicate the covariant derivative for the scalar triplet $\Phi(x)$
in (\ref{covariantderivforPhi}) by $\cD_\mu(A, \bA)$ to show its dependence
on the gauge fields. Then the gauge-fixings for the H and EW symmetries
transform it as follows:
\begin{equation}
   \cD_\mu(A^{\pr}, \bA^{\pr})=
    \Omega^{-1}(\bvth(x))\Omega_{\rm EW}^{-1}(\vartheta(x))
    \cD_\mu(A, \bA)\Omega_{\rm EW}(\vartheta(x)) \Omega(\bvth(x))
       \label{HFEWUnitaryPhaseofA}
\end{equation}
where $A_\mu^{\pr}(x)$ is the vector fields in the phase of broken symmetries.
To derive configurations of the transformed vector fields $A_\mu^\pr(x)$ and
examine their interaction with the scalar component fields, it is necessary
to investigate the action of this covariant derivative on the scalar triplet
$\Phi_0(x)$ in (\ref{Phidecomposition}). Note that, in such calculation,
effects of the super-massive vector fields $\bW_\mu(x)$, $\bY_\mu(x)$ and
$\bZ_\mu(x)$ can safely be ignored. To determine the configurations
of the vector fields $A_\mu^{\pr}(x)$, it is sufficient to calculate the
action of the covariant derivative $\cD_\mu(A^{\pr}, 0)$ on the triplet
$\Phi_0(x)$. Consequently, we can reproduce all of the results of the
Weinberg-Salam theory and determine interactions of the scalar fields
$\zeta^{+}_i(x)\,(i=1,\,2)$, $\zeta^0_1(x)$ and $\eta_i(x)\,(i=1,\,2,\,3)$
with the electromagnetic field $A_\mu(x)$ and the weak boson fields
$W_\mu(x)$ and $Z_\mu(x)$ which possess, respectively, the masses 
$M_{W}^2 = \half g_2^2v^2$ and $M_{Z}^2 = \half (g_2^2+g_1^2)v^2$.

In total, the triplet $\Phi(x)$ is proved to break the H and EW symmetries
at the scale $\Lambda$ and to create the weak boson fields as well as the
electromagnetic field from the EW gauge fields without leaving any massless
scalar fields at all. Here, the transferring of freedoms of the scalar fields
to the gauge fields is traced as follows:
\begin{equation}
\begin{array}{c}
 \left\{\ 
  \begin{array}{cl}
   2 {\rm \ gauge\ fields\ } A^{(2)1}_\mu(x) {\rm \ and\ }
   A^{(2)2}_\mu(x) & :\ 2\times 2 \\
   \noalign{\vskip 0.1cm}
   3 {\rm \ massless\ complex\ scalar\ fields\ } \phi_{i}^{+}(x)\,
    (i=1,\,2,\,3) & :\ 3\times 2
  \end{array}\ 
 \right\} \\
  \noalign{\vskip 0.1cm}
   \Downarrow \\
  \noalign{\vskip 0.1cm}
 \left\{\ 
  \begin{array}{cl}
   1 {\rm \ massive\ charged\ vector\ field\ } W^{+}_\mu(x)
   & :\ 2\times 3 \\
  \noalign{\vskip 0.1cm}
   2 {\rm \ massive\ complex\ scalar\ fields\ } \zeta_{1}^{+}(x)
     {\rm \ and\ } \zeta_{2}^{+}(x)
   & :\ 2\times 2 \\
  \end{array}\ 
 \right\}
\end{array}
\label{BalancePhiUp}
\end{equation}
for the EW up-sector, and
\begin{equation}
\begin{array}{c}
 \left\{\ 
  \begin{array}{cl}
   2 {\rm \ gauge\ fields\ } A^{(2)3}_\mu(x) {\rm \ and\ } A^{(1)}_\mu(x)
   & :\ 2\times 2 \\
  \noalign{\vskip 0.1cm}
   3 {\rm \ massless\ complex\ scalar\ fields\ } \phi_{i}^{0}(x)\,
    (i=1,\,2,\,3) 
   & :\ 3\times 2 \\ 
  \end{array}
 \right\} \\
  \noalign{\vskip 0.1cm}
   \Downarrow \\
  \noalign{\vskip 0.1cm}
 \left\{\ 
  \begin{array}{cl}
   1 {\rm \ massive\ real\ vector\ field\ } Z_\mu(x)
   & :\ 1\times 3 \\
   1 {\rm \ massless\ electromagnetic\ field\ } A_\mu(x)
   & :\ 1\times 2 \\
  \noalign{\vskip 0.1cm}
   1 {\rm \ massive\ complex\ scalar\ field\ } \zeta_{1}^{0}(x)
   & :\ 1\times 2 \\
   3 {\rm \ massive\ real\ scalar\ fields\ } \eta_{i}(x)\, (i=1,\,2,\,3)
   & :\ 3\times 1\\
  \end{array}\ 
 \right\}
\end{array}
\label{BalancePhiDown}
\end{equation}
for the EW down-sector. Preservation of the number of the
independent modes of the fields is confirmed, respectively, as
$2\times 2 + 3\times 2 = 10 = 2\times 3 + 2\times 2$ for the up-sector and
$2\times 2 + 3\times 2 = 10 = 1\times 3 + 1\times 2 + 1\times 2 + 3\times 1$
for the down-sector, before and after the phase transition.

%\subsection{Dirac mass matrices for quarks and leptons}
At the high energy scale $\brLam$, the unitary gauge factor is
separated out of the fermion triplets leaving the skeleton triplets
$\uPsi^{\,f}_h(x)$ in (\ref{HFUnitaryPhaseofPhiPsi}). In the low energy
region around the scale $\Lambda$,  the renormalization group effects
are presumed to be properly taken into account for all of the quantities
in the Lagrangian densities for the Yukawa interactions, (\ref{YukawaUp})
and (\ref{YukawaDown}).
%For the sake of simplicity, the same symbols will be used for
%all quantities in the Lagrangian.

Then, through the breakdown of EW symmetry at the scale $\Lambda$, the
fermion fields acquire masses of Dirac type. Substitution of the decomposition
of $\Phi_0(x)$ in (\ref{Phidecomposition}) into (\ref{YukawaUp})
and (\ref{YukawaDown}) leads to the effective Lagrangian density
for the fermion fields in the low energy region as
\begin{equation}
 \cL_{\rm Y}\ \rightarrow\ 
 \cL^{\rm Y}_{\cM}
 = \sum_{f=u,d,\nu,e}\,\bar{\uPsi}^{\,f}_{L0}\cM_f\uPsi^{\,f}_{R0}
  + {\rm h.c.} + \cdots
\end{equation}
where $\uPsi^{\,f}_{h0}=\Omega_{\rm EW}(\th)^{-1}\uPsi^{\,f}_h$, and $\cM_f$
are the mass matrices of Dirac type. The ellipsis stands for the
interactions of fermion and scalar fields. For the up-sectors ($f=u,\,\nu$)
of EW symmetry, we deduce the Dirac mass matrices as follows:
\begin{equation}
   \cM_{f} = a_f\,I
              + \frac{1}{3}b_{f1}\left(
                       \begin{array}{ccc}
                         -1 & -1 & 2 \\
                         -1 & -1 & 2 \\
                         -1 & -1 & 2 \\
                       \end{array}
                     \right)
             + \frac{1}{\sqrt{3}}
               b_{f2}\left(
                      \begin{array}{rrr}
                         1 &  1 &  1  \\
                        -1 & -1 & -1 \\
                         0 &  0 &  0 \\
                      \end{array}
                     \right)+ c_f\,\bD
   \label{EWUpDirac}
\end{equation}
where $a_f = Y_{f3}\,v$, $b_{f1} = Y_{f2}\,v$, $\ b_{f2} = -Y_{f1}\,v$
and $c_f = 3Y_{f4}v$. 
Likewise, for the down-sector ($f=d,\,e$) of EW symmetry, we obtain
\begin{equation}
   \cM_{f} = a_f\,I
             + b_{f1}\left(
                      \begin{array}{ccc}
                        0 & 0 & 0 \\
                        0 & 0 & 0 \\
                        1 & 1 & 1 \\
                      \end{array}
                     \right)
             + \frac{1}{\sqrt{3}}
                b_{f2}\left(
                       \begin{array}{ccc}
                         1 & -1 & 0 \\
                         1 & -1 & 0 \\
                         1 & -1 & 0 \\
                       \end{array}
                     \right)
             + c_f\,\bD
    \label{EWDownDirac}
\end{equation}
where $a_f = Y_{f3}\,v$, $\ b_{f1} = Y_{f1}\,v$, $\ b_{f2} = Y_{f2}\,v$
and $c_f = 3Y_{f4}\,v$. In each sector, the number of the free parameters
is the same with that of the Yukawa coupling constants.

Apparently, these rather peculiar mass matrices are not self-adjoint.
To diagonalize such matrices, it is necessary to resort to the
bi-unitary transformation~\cite{Chang Li}
\begin{equation}
    V_L^{f\dag}{\cal{M}}_{f} V_R^f = {\cal M}_{\rm diagonal} .
\end{equation}
To derive mass eigenvalues, we have to solve the eigenvalue problem
for the self-adjoint matrices $\cM_f\cM^{\dag}_f$
as follows:
\begin{equation}
 \cM_f\cM^{\dag}_f |\boldv^{(f)i}\rangle = m_{(f)i}{}^2\,|\boldv^{(f)i}\rangle .
  \label{mass2eigenvalue}
\end{equation}
The diagonalizing matrix $V_L^{f}$ is obtained from the eigenvectors.
%\begin{equation}
%    V_L^f = \left(\,|\boldv^{(f)1}{}\rangle,\ |\boldv^{(f)2}\rangle,\ 
%    |\boldv^{(f)3}\rangle \,\right).
%\end{equation}
For the charged fermion families ($f=u,\,d$ and $e$), solutions of this
eigenvalue problems are sufficient to obtain information on mass spectra
and diagonalizing matrices. The FMM for quark sector is constructed
in the form $V=V_L^{u\dag}V_L^{d}
=\left(\,\langle \boldv^{(u)i}|\boldv^{(d)j}\rangle\,\right)$. 

\section{Discussion}
We have developed the gauge field theory of the V and H symmetries
which brings forth an effective theory for unified description of
flavor physics in low energy regime. The H symmetry generated by
the central extension of the Pauli algebra works to reduce uncertainty
and create orders in the Yukawa interactions. In particular,
the H-sum of the triplet plays essential roles to provide
unique structures to the Yukawa interactions.

To make the effective theory to be {\lq\lq}effective{\rq\rq} in low energy
flavor physics, the number of its unknown parameters should be restricted
as less as possible. For its purpose, we have formulated the new scheme of
the compound mechanism of symmetry breakdown in which the reference state
for partially broken symmetry is chosen by specifying the single parameter
for the stationary point of the Higgs potential and the freedoms arising in
the decomposition of the scalar triplet around the reference state are used
to complete the symmetry breakdown and to forbid unphysical modes to appear.

At the high energy scale $\brLam$, we determined the reference state
for the partially broken H symmetry by calculating the stationary-point of
the Higgs potential $V_1(\bPhi)$ and made use of the mixing effect between
the component fields of the decomposition of $\bPhi(x)$ to exclude the NG
mode. At the low energy scale $\Lambda$, we had to replace
the pseud-reference state which is fixed by the stationary-point of the part
of the Higgs potentials $V_2(\Phi)+V_3(\Phi, \bv)$ with the reference state
by rescaling the parameter from $v_0$ to $v$ and utilized the mixing between
the component fields of the decomposition of $\Phi(x)$ to suppress both of
the NG and tachyonic modes.

The effective theory brought forth by the compound mechanism of symmetry
breakdown provides the mass matrices of Dirac type for each sector of
basic fermions and the mass matrix of Majorana type for the right-handed
chiral neutrinos in the unified way. All of the mass matrices have hierarchical structures. The compound mechanism which is designed not to increase the
number of free parameters works successfully to restrict the Dirac mass matrix
$\cM_f$ of $f$-sector to possess only four complex numbers. 
%$a_f$, $b_{f1}$, $b_{f2}$ and $c_f$
%In actual anlysis of flavor physics, it is not $\cM_f$ but $\cM_f\cM^{\dag}_f$
Here it is worth to recognize that multiplication rules of component
matrices in each $\cM_f$ can reduce the number of independent parameters
in $\cM_f\cM^{\dag}_f$. Direct calculation proves that $\cM_f\cM^{\dag}_f$
can be expressed in terms of four real numbers and two phase variables.

In the eigenvalue problems in (\ref{mass2eigenvalue}) for the quark
sector, the eigenvalues for squared masses of the up and down quarks are
specified by six real parameters in the mass matrices $\cM_u$ and $\cM_d$.
Then, we are able to express the FMM composed of the eigenvectors in terms
of ten parameters: six eigenvalues, two remaining real parameters and
two phase differences. On the other hand, ten different kinds of reliable
experimental data, i.e., six mass values and four parameters in the FMM are
available in the compilation by Particle Data Group.~\cite{PDG} Therefore,
although no prediction can be made, we are able to apply the present scheme
to the data analyses to verify its consistency and to explore hitherto
unknown features of quark flavor.
Physical situations are much involved for the lepton sector, since
the neutrinos have also the Majorana masses. Recent observations of neutrino
oscillations have confirmed that neutrino family possesses minute but
non-vanishing squared mass differences.\cite{solar,atmospheric,KamLAND}
One promising way to account for smallness of the neutrino masses is
the seesaw mechanism.%~\cite{Gell-Mann,Yanagida}
To determine the neutrino mass spectrum and the FMM for the lepton
sector, we must solve the eigenvalue problem for a $6\times 6$ matrix
consisting of the Dirac mass matrix $\cM_\nu$ in (\ref{EWUpDirac})
and the Majorana mass matrix $\breve{\cM}_\nu$ in (\ref{Majoranamass}). 
Analyses of the mass matrices for quark and lepton sectors will be made
in future investigation.

Our theory predicts the existence of rich physical modes of bose fields.
Two and six kinds of bose fields descend, respectively, from the triplets
$\bPhi(x)$ and $\Phi(x)$. At the present stage, it is impossible to fix
their masses theoretically. It is important, however, to recognize that
some of their basic characteristics do not depend on details of the scheme
of symmetry breaking. It is, in principle, the representations of
the EW and H symmetries that determine patterns of the interactions of
the scalar fields with themselves and with other fields.
The number of those fields is determined, as shown in the balance sheets
in (\ref{BalancebPhi}), (\ref{BalancePhiUp}) and (\ref{BalancePhiDown}),
by the residual degrees of the scalar triplets which are not transferred
to the gauge fields.

The real scalar field $\bxi_1(x)$ and $\bxi_2(x)$ descendant from the scalar
H triplet $\bPhi(x)$ do not interact with fundamental fermions except for
the neutrinos. From (\ref{VEVbv}) and (\ref{massbxi1}), the squared mass of
the would-be NG boson field is derived to be
\begin{equation}
    m_{\tbxi_1}^2 = \frac{(\brlam_1+\brlam_3)\bm_2^2
                          -(\brlam_2+\brlam_3)\bm_1^2}
                       {\brlam_1+\brlam_2+2\brlam_3} \propto \sin\bth .
\end{equation}
Note that, although it might appear paradoxical, there exist
a possibility for the mass $m_{\tbxi_1}$ of this boson created around
the high energy scale $\brLam$ to be comparable to or less than those
of the bosonic fields produced around the low energy scale $\Lambda$.
If this neutral field has the smallest mass among massive boson fields,
it may survive a long life through evolution of the universe and give
essential influence necessarily on the history of the universe.
%Results of the cosmic ray observatory such as PAMELA~\cite{PAMERA} are
%expected to provide information which clarify detailed nature of dark matter.
It is suggestive to interpret the fields $\bxi_1(x)$ and $\bxi_2(x)$ which
are blind to the SM quantum numbers might be related with the dark matter
and also the dark energy being responsible for the late-time accelerating
expansion of the universe.\cite{Riess1,Perlmutter,Riess2,WMAP}

Nine remaining degrees of component fields of the scalar triplet $\Phi(x)$
are survived as six kinds of massive scalar fields: three complex fields
$\zeta_1^{+}(x)$, $\zeta_2^{+}(x)$ and $\zeta_1^0(x)$; three real fields
$\eta_j(x)$ $(j=1,\,2,\,3)$.  The real field $\eta_3(x)$ is just the one
corresponding to the Higgs boson in the Weinberg-Salam (WS) theory and
all of the remainders are new fields predicted in our theory. It is
instructive to consider the case where no rescaling of the reference state
is made and no mixing effect exist, i.e., $v=v_0$ ($Z=1$) and $\theta=0$.
In such a limiting case, two fields $\eta_1(x)$ and $\eta_2(x)$
become massless. Therefore, both of the rescaling and mixing effects are
necessary to make those fields to be massive.

%At the present stage, however, we are not
%able to foretell values of masses of these seven scalar particles,
%since the theory possesses too many parameters. 
Note here that there exist several natural constraints on unknown coupling
constants. For the Higgs mechanism can be realized at the low energy scale,
the squares of effective Higgs masses, $m^{\pr 2}_1$ and $m^{\pr 2}_2$ in
(\ref{effectiveHiggsmass}), and the parameter $v_0^2$ in (\ref{v0square})
must be positive. These conditions require that the combinations of the coupling
constants of mutual-interaction between the triplets $\Phi(x)$ and $\bPhi(x)$,
$\dlam_1+\dlam_2$, $\dlam_3+\dlam_4$ and $(4\dlam_1+\dlam_2)/9$ must take
smaller values than $(m_1^2+m_2^2)/\bv^2$. Contrastingly, it is favorable for
those mutual coupling constants $\dlam_j$ $(j=1,\,\cdots,\,7)$ to have some
lower bounds to guarantee that the squared masses in (\ref{massscalarlow1})
and (\ref{massscalarlow2}) are positive and take large values. However, such
adjustment of $\dlam_j$ is not sufficient to make those squared masses to be
positive definite. The relation in (\ref{Zcondition}), being independent of
the constant $\dlam_j$, shows that the coupling constants $\balam_j$
($j=1,\cdots\,4$) and $\wtlam_3$ must also be properly chosen.

The six bose fields can interact with all kinds of the basic fermion fields.
However, the experiment establish that the fermions are highly suppressed
to communicate with each other through exchanges of neutral scalar fields.
This feature known as suppression of the flavor changing neutral current
(FCNC) imposes very strong constraints on the masses of the fields $\eta_j(x)$
$(j=1,\,2,\,3)$ and $\zeta_1^0(x)$. The FCNC bound which results from the
$B-\bar{B}$ and $K-\bar{K}$ mixings requires their masses to be larger than
$10^2 \sim 10^3 {\rm TeV}$.~\cite{FCNC} To see implications of this constraint,
let us examine the relation in (\ref{Zcondition}). Assuming roughly that
the constants $\balam_j$ and $\wtlam_3$ are almost of the same order and that
$Z\sim 10$, we find $\balam v^2 \geq 10^2\sim 10^4 {\rm TeV}^2$. As is well
known, from the mass of the weak charged boson $W_\mu^{+}(x)$ and the Fermi
coupling constant $G_{\rm F}$, the EW scale parameter $v$ can be estimated
by $v^2\simeq 1/(2\sqrt{2}G_{\rm F}) \approx 4\times 10^4 {\rm GeV}^2$.
Therefore, to satisfy the FCNC constraint, the coupling constants $\balam_j$
and $\wtlam_3$ must be larger than $10^4 \sim 10^6$.
Compared with the FCNC, the constraints of the flavor changing charged currents
are not so stringent. From (\ref{massscalarlow1}) and (\ref{massscalarlow2}),
it is unnatural to regard that the charged bosons $\zeta_1^{+}(x)$ and
$\zeta_2^{+}(x)$ possess the masses of same order with the neutral bosons.
It is attractive to consider that a rich spectrum of the scalar fields
$\bxi_j(x)$, $\zeta_j^+(x)$, $\zeta_1^0(x)$ and $\eta_j(x)$ could be
possible targets of high energy experiments by the LHC group in future. 

At this stage, our theory should be evaluated as a provisional scheme of
gauge field theory which is designed to result in one effective theory for
flavor physics. The symmetries are broken in a special way, called the compound
mechanism, which utilizes the freedoms arising in the choice of the reference
state and in the decomposition of the scalar triplet around it. Due efforts
must be made to justify the compound mechanism of symmetry breakdown in the
context of the quantum field theory. We must find physical meanings of the
rescaling of the reference state and of the constant $Z(>1)$.

We have chosen the simple form, in (\ref{VEVbPhi}) and (\ref{VEVPhi}),
for the reference state of partially broken symmetry specified by a single
parameter, not to increase the numbers of unknown parameters in the
low-energy effective theory. As an extension of our theory, we must examine
the reference state which is specified by many parameters and breaks the
symmetry totally without any additional assumption. The theory of the private
Higgs proposed by Porto and Zee~\cite{Zee} is an example of many parameters.
There may exist such an economical scheme with multi-parameters that can
produce a low energy effective theory with handy mass matrices.

%It is necessary to continue our efforts to formulate an extended
%unified gauge field theory for the V$\times$H symmetry. 
The conventional grand unified scheme%\cite{GeorgiGlashow,FritzschMinkowski}
is framed, independently of the H symmetry, by unifying the vertical symmetry
into a simple group. In our scheme, however, the H symmetry
is incorporated into the gauge field theory in a close connection with
the EW symmetry. Therefore, we must find a new path to combine the color
symmetry with the EW$\times$H symmetry to formulate an extended grand unified
theory for particle physics. To avoid the difficulty of quadratic divergence
and solve the hierarchy problem, it is requisite to challenge
the super-symmetric generalization of the present theory and its extended grand
unified version.

\section{Acknowledgement}
I would like to express my sincere thanks to Professor Y. Koide and
Professor K. Yamawaki for their valuable comments and encouragement.

\appendix

\section{EW$\times$H invariants of scalar fields}
In the WS theory, the Higgs potential of the EW doublet $\phi(x)$ possesses
the compact form composed solely of the invariant $\phi^{\dag}\phi$. This
is because the internal space of the EW isospin $\{\,\tau_a\,\}$
has simple structure. For $\phi$, the dual field can be defined
by $\tilde{\phi}=i\tau_2\phi^\ast$ so that $\tilde{\tilde{\phi}}=\phi$.
These fields which have the same transformation property under
the EW symmetry group satisfy the relation $\phi^{\dag}\tilde{\phi} = 0$.
The {\lq\lq}internal vector{\rq\rq} $\phi^{\dag}\tau_a\phi$ satisfy a simple
identical equation, i.e., its squared length can be expressed in terms of the
pure {\lq\lq}internal scalar{\rq\rq} quantity as follows:
\begin{equation}
  (\phi^{\dag}\tau_a\phi)(\phi^{\dag}\tau^a\phi) = (\phi^{\dag}\phi)^2 .
  \label{EWsimplerelation}
\end{equation}

In contrast to the WS theory, the internal space of EW$\times$H symmetry
has complex structure. First of all, it is impossible to define
the {\em dual} fields for the triplets $\bPhi(x)$ and $\Phi(x)$.
For example, the associated field for $\wtbPhi(x)$ defined
in (\ref{AssociatebPhi}) can not be {\em dual} to $\bPhi(x)$, since
\begin{equation}
 \widetilde{\wtbPhi}(x) = -(I-\bD)\bPhi \neq \bPhi.
\end{equation}
Nevertheless, it is possible to prove that, as in (\ref{EWsimplerelation})
in the SM, all of the quartic invariants composed of internal vectors and
bi-vectors are expressed in terms of the pure {\lq\lq}internal scalar{\rq\rq}
quantities. The proof is given below in three steps.

%For construction of the Lagrangian density of the scalar triplets,
%we must examine all of the invariants and find out relations among them.
Some of the identical equations can be proved directly by the definitions.
For example, the H-sums of the associated scalar triplets $\wtbPhi(x)$
and $\wtPhi(x)$ vanish, i.e.,
\begin{equation}
     \ok \wtbPhi \ck = 0,\quad \ok \wtPhi \ck = 0
\end{equation}
due to the action of $\btau_2$.
To examine and classify the invariants in general, however, it is necessary
to use explicit representations of the scalar triplets in terms of the
eigenvectors in (\ref{eigenvectors}):
\begin{equation}
  \bPhi = z_1|\,1\,\rangle +z_2|\,2\,\rangle +z_3|\,3\,\rangle,
  \quad
  \wtbPhi = z_2^\ast|\,1\,\rangle -z_1^\ast|\,2\,\rangle
  \label{ExplicitbPhiDecomposition}
\end{equation}
where $z_j(x)$ are complex scalar fields, and
\begin{equation}
  \Phi = Z_1|\,1\,\rangle +Z_2|\,2\,\rangle +Z_3|\,3\,\rangle,
  \quad
  \wtPhi = i\tau_2 Z_2^\ast|\,1\,\rangle -i\tau_2 Z_1^\ast|\,2\,\rangle
  \label{ExplicitPhiDecomposition}
\end{equation}
where $Z_j(x)={}^t(\zeta_j^+,\,\zeta_j^0)$ are complex scalar fields
of EW doublets.

Using the decompositions in (\ref{ExplicitbPhiDecomposition}), the triplets
$\bPhi(x)$ and $\wtbPhi(x)$ are proved to satisfy the identity
\begin{equation}
    \bPhi^{\dag}\wtbPhi = 0 .
    \label{makebPhisimple}
\end{equation}
It is this identity that simplifies the quartic relations of the triplet
$\bPhi$. The bilinear form of the associated triplet $\wtbPhi$ is reduced
to those of the triplet $\bPhi$ as
\begin{equation}
    \wtbPhi^{\dag}\wtbPhi
       = \bPhi^{\dag}(I-\bD)\bPhi
       = \bPhi^{\dag}\bPhi - \frac{1}{3}\ok\bPhi^{\dag}\ck\ok\bPhi\ck .
\end{equation}
Quartic relations of the triplets satisfy the following identical equations as
\begin{equation}
    \left(\bPhi^{\dag}\btau_j\wtbPhi\right)
    \left(\wtbPhi^{\dag}\btau^j\bPhi\right)
     = 2\,\left(\bPhi^{\dag}(I-\bD)\bPhi\right)^2
\end{equation}
and
\begin{equation}
    \left(\bPhi^{\dag}\btau_j\bPhi\right)
    \left(\bPhi^{\dag}\btau^j\bPhi\right)
     = \left(\bPhi^{\dag}(I-\bD)\bPhi\right)^2
     = \left(\bPhi^{\dag}\bPhi
     - \frac{1}{3}\ok\bPhi^{\dag}\ck\ok\bPhi\ck\right)^2 .
\end{equation}
These quartic relations show that invariant combinations including internal
vectors composed of scalar triplet-spinors are reducible to invariant quantities
consisting of $\bPhi^{\dag}\bPhi$ and $\ok\bPhi^{\dag}\ck\ok\bPhi\ck$.

In contrast with the identity in (\ref{makebPhisimple}), $\Phi(x)$ and
$\wtPhi(x)$ are {\em not orthogonal} with each other. The decompositions in
(\ref{ExplicitPhiDecomposition}) leads readily to
\begin{equation}
    \Phi^{\dag}\wtPhi
     = Z_1^\dag i\tau_2 Z_2^\ast - Z_2^\dag i\tau_2 Z_1^\ast
     = 2(\zeta_1^-\zeta_2^{0\ast}-\zeta_2^-\zeta_1^{0\ast}) \neq 0 .
    \label{PhiTildePhi}
\end{equation}
This turns out to be a main cause of complexity of the Higgs potential
$V_2(\Phi)$ in (\ref{phipot}). The triplets $\Phi(x)$ and $\wtPhi(x)$
satisfy the following quadratic relations as
\begin{equation}
    \wtPhi^{\dag}\wtPhi = {}^{\bt}\Phi(I-\bD)\Phi^\ast
    = \Phi^{\dag}\Phi - \frac{1}{3}\ok\Phi^{\dag}\ck\ok\Phi\ck ,
\end{equation}
\begin{equation}
    \Phi^{\dag}\btau_j\wtPhi = 0, \quad
    \Phi^{\dag}\tau_a\wtPhi = 0, 
\end{equation}
\begin{equation}
    \wtPhi^{\dag}\btau_j\wtPhi = - \Phi^{\dag}\btau_j\Phi, \quad
    \wtPhi^{\dag}\tau_a\wtPhi =  - \Phi^{\dag}\tau_a\Phi . 
\end{equation}

By using the decomposition in (\ref{ExplicitPhiDecomposition}), we are
able to prove the following quartic relations of the scalar triplets as
\begin{equation}
 \left(\Phi^{\dag}\btau_j\Phi\right)\left(\Phi^{\dag}\btau^j\Phi\right)
 = |\Phi^{\dag}(I-\bD)\Phi|^2 - |\Phi^{\dag}\wtPhi|^2 ,
\end{equation}
\begin{equation}
 \left(\Phi^{\dag}\tau_a\Phi\right)\left(\Phi^{\dag}\tau^a\Phi\right)
 = |\Phi^{\dag}\Phi|^2
 + 2\Phi^{\dag}i\tau_2{}^{\bt}\Phi^{\ast}{}^t\Phi i\tau_2\Phi,
\end{equation}
\begin{equation}
 \left(\Phi^{\dag}\tau_a(I-\bD)\Phi\right)
  \left(\Phi^{\dag}\tau^a(I-\bD)\Phi\right)
  = |\Phi^{\dag}(I-\bD)\Phi|^2 - |\Phi^{\dag}\wtPhi|^2 ,
\end{equation}
\begin{equation}
 \left(\Phi^{\dag}\tau_a\bD\Phi\right)\left(\Phi^{\dag}\tau^a\bD\Phi\right)
 = |\Phi^{\dag}\bD\Phi|^2
 = \frac{1}{9}\left(\ok\Phi^{\dag}\ck\ok\Phi\ck\right)^2 ,
\end{equation}
\begin{equation}
     \left(\Phi^{\dag}\tau_a\btau_j\Phi\right)
     \left(\Phi^{\dag}\tau^a\btau^j\Phi\right)
     = |\Phi^{\dag}(I-\bD)\Phi|^2 + 2|\Phi^{\dag}\wtPhi|^2 ,
\end{equation}
\begin{equation}
     \left(\Phi^{\dag}\tau_a\btau_j\wtPhi\right)
     \left(\wtPhi^{\dag}\tau^a\btau^j\Phi\right)
     = 4|\Phi^{\dag}(I-\bD)\Phi|^2 - |\Phi^{\dag}\wtPhi|^2 .
\end{equation}
With these identical relations, all invariants composed of internal vectors
and bi-vectors composed of the triplets of EW doublets can be reduced to
the {\lq\lq}internal scalar{\rq\rq} invariants composed.

There are two identical relations concerning the internal vectors of
the triplets $\bPhi(x)$ and $\Phi(x)$ as
\begin{equation}
  (\bPhi^\dag\btau_j\Phi)^\dag(\bPhi^\dag\btau^j\Phi)
  = (\bPhi^\dag(I-\bD)\bPhi)(\Phi^\dag(I-\bD)\Phi)
  + (\Phi^\dag\wtbPhi)^\dag(\Phi^{\dag}\wtbPhi)
\end{equation}
and
\begin{equation}
  (\bPhi^\dag\btau_j\bPhi)^\dag(\Phi^\dag\btau^j\Phi)
  = (\bPhi^\dag(I-\bD)\bPhi)(\Phi^\dag(I-\bD)\Phi)
  - 2(\Phi^\dag\wtbPhi)^\dag(\Phi^{\dag}\wtbPhi) .
\end{equation}
Evidently, the triplets $\bPhi(x)$ and $\Phi(x)$ possess quartic
invariants $|\Phi^\dag(I-\bD)\bPhi|^2$ and $|\Phi^\dag\bD\bPhi|^2$.
Further, there exist intricate terms such as
$(\Phi^\dag\bPhi)(\bPhi^\dag\Phi)$,
$(\Phi^\dag\bPhi)(\ok\bPhi\ck^\dag\ok\Phi\ck)$,
$(\bPhi^\dag\Phi)(\ok\Phi\ck^\dag\ok\bPhi\ck)$ and
$(\wtPhi^\dag\bPhi)(\bPhi^{\dag}\wtPhi)$ =
$(\Phi^\dag\wtbPhi)(\wtbPhi^{\dag}\Phi)$. However, the conservation of
the H hypercharge and the assignment of $\by_{\scriptsize \bPhi} \neq 0$
and $\by_{\Phi} = 0$ prohibit these terms, except for the last one,
to appear in the Higgs potential $V_3(\Phi,\,\bPhi)$.

With resort to the identical relations proved in this Appendix, it turns
out possible to express all EW$\times$H invariants in terms of the internal
{\lq\lq}scalar{\rq\rq} quantities.

%\begin{equation}
%     \tan\bth = \sqrt{2}
%     \frac{\bm_1^2-(\brlam_1+\brlam_3)\bv^2}
%     {\bm_1^2+3\bm_2^2-(\brlam_1+3\brlam_2+4\brlam_3)\bv^2}
%      \equiv \sqrt{2}\br
%\end{equation}
%and, consequently, determines the parameters as
%\begin{equation}
%    \balph=\cos\bth = \frac{1}{\sqrt{1+2\br^2}},\ \ 
%    \bbeta=\sin\bth = \frac{\sqrt{2}\br}{\sqrt{1+2\br^2}} .
%\end{equation}
%\begin{equation}
%   m_{\tbxi 1}^2 = -\bm_1^2 + (\brlam_1+\brlam_3)\bv^2
%\end{equation}
%and
%\begin{equation}
%  \begin{array}{l}
%   m_{\tbxi 2}^2 = m_{\tbxi 1}^2 
%   -\frac{6\br^2}{1+2\br^2}\bm_2^2\\
%   \noalign{\vskip 0.2cm}
%   \qquad\quad
%   +\frac{2}{3(1+2\br^2)}\left[2\brlam_1-4(\brlam_1+3\brlam_3)\br
%             +(2\brlam_1+27\brlam_2+21\brlam_3)\br^2\right]\,\bv^2
%  \end{array}
%\end{equation}

\end{document}